\documentclass[aps,prd,groupedaddress]{revtex4}
\usepackage{graphicx}
\usepackage{epsfig}
\usepackage{amsmath}
\usepackage{amsfonts}
\usepackage{amssymb}
\usepackage{epsf}
\newcommand{\insertplot}[5]{\begin{figure}
 \hfill\hbox to 0.05in{\vbox to #5in{\vfill
 \inputplot{#1}{#4}{#5}}\hfill}
 \hfill\vspace{-.1in}
 \caption{#2}\label{#3}
 \end{figure}}
\newcommand{\inputplot}[3]{
 \special{ps: plotfile #1}
}

\usepackage{cancel}
\usepackage{ulem}
\usepackage{xcolor}

\newcounter{fig}

\newcommand{\atan}{\;{\rm arctan}}
\newcommand{\hl}{\hat{l}}

\begin{document}
\title{Quasi-periodic Oscillations in Rotating Ellis Wormhole Spacetimes}
\author{Efthimia Deligianni}
\email[]{efthimia.deligianni@uni-oldenburg.de}
\affiliation{Institute of Physics, University of Oldenburg, D-26111 Oldenburg, Germany}
\author{Burkhard Kleihaus}
\email[]{b.kleihaus@uni-oldenburg.de}
\affiliation{Institute of Physics, University of Oldenburg, D-26111 Oldenburg, Germany}
\author{Jutta Kunz}
\email[]{jutta.kunz@uni-oldenburg.de}
\affiliation{Institute of Physics, University of Oldenburg, D-26111 Oldenburg, Germany}
\author{Petya Nedkova}
\email[]{pnedkova@phys.uni-sofia.bg}
\affiliation{Faculty of Physics, Sofia University, Sofia 1164, Bulgaria}
\author{Stoytcho  Yazadjiev}
\email[]{yazad@phys.uni-sofia.bg}
\affiliation{Faculty of Physics, Sofia University, Sofia 1164, Bulgaria\\
Bulgarian Academy of Sciences, Sofia 1113, Bulgaria}

\date{\today}
\begin{abstract}
We analyze the properties of the circular orbits for massive particles in the equatorial plane
of symmetric rotating Ellis wormholes.
In particular, we obtain the orbital frequencies and the radial and vertical epicyclic frequencies,
and consider their lowest parametric, forced and Keplerian resonances.
These show that quasi-periodic oscillations
in accretion disks around symmetric rotating Ellis wormholes have many distinct properties
as compared to quasi-periodic oscillations
in accretion disks around rotating Teo wormholes and the Kerr black hole. Still we can distinguish some common features which appear in wormhole spacetimes as opposed  to black holes. The most significant ones include the possibility of excitation of stronger resonances such as lower order parametric and forced resonances and the localization of these resonances deep in the region of strong gravitational interaction near the wormhole throat, which will lead to further amplification of the signal.
\end{abstract}

\maketitle

\section{Introduction}

General relativity is well tested in the weak field regime. However, its predictions for strong gravitational fields  are still under scrutiny. Current and near future observational missions will bring insights about the precision with which general relativity describes the high energy processes in our universe narrowing down the options for introducing modifications. Another emerging issue is to explore the full variety of  compact objects, which are astrophysically justified. Besides the well accepted neutron stars and black holes, the gravitational theories predict more exotic self-gravitating systems like wormholes and naked singularities which  might be detected experimentally.

A promising channel for probing the strong gravitational interaction is the X-ray spectroscopy. The accreting matter in the vicinity of the compact objects emits X-ray radiation, the major part of which is produced in the innermost part of the accretion disk at the range of several gravitational radii. Thus, it contains information about some fundamental relativistic properties of the gravitational field. Several features of the X-ray spectrum are analysed in this regard like the temperature of the soft X-ray continuum emission or the profile of the iron K$\alpha$ line, which give estimates about the position of the ISCO and the angular momentum of the central compact object. However, the best experimental precision is expected from the measurements of the quasi-periodic oscillations (QPOs) in the accretion flux. If they are interpreted within the correct physical model, the QPOs can provide a potentially more reliable method for determining the spin of the compact object. 

The QPOs are peaks in the variability of the X-ray energy occurring at some characteristic frequency ratio  for a wide range of compact objects in different physical conditions like neutron stars, stellar mass black holes or active galactic nuclei. For microquasars mostly two peaks are observed in the kHz range with frequencies scaling as $3:2$. At the moment sufficiently reliable data is available only for a few astrophysical sources but the  upcoming generation of spectrometers like eXPT \cite{Zhang:2016}, LOFT  \cite{Feroci:2016} ans STROBE-X \cite{Wilson-Hodge:2017} are expected to bring diversity and  precision.

The exact physical processes which lead to the formation of the QPOs in the accretion flow are currently not completely understood but their observational behavior gives strong indications that they are a hydrodynamical phenomenon occurring in the relativistic gravitational field. Viable suggestions for their explanation include the disk seismological models in which the QPOs are associated with different trapped modes in the disk oscillations \cite{Kato:1980}-\cite{Rezzolla}. Another approach approximates the disk eigenfrequencies with the characteristic frequencies of the oscillating circular geodesics in the equatorial plane and searches for a relation between combinations of these frequencies and the QPOs. In this direction the relativistic precession model identifies the twin peak kHz QPOs with the geodesic orbital frequency and the periastron precession frequency \cite{Stella:1999}. On the other hand the resonance model suggests that the QPOs arise due to the excitation of certain non-linear resonances in the orbit oscillations and inherit the corresponding resonance frequencies \cite{Abramowicz:2001}-\cite{Torok:2005a}. In the linear approximation the circular orbit oscillations are described by two eigenfrequencies corresponding to the radial and vertical epicyclic frequencies. When non-linear effects are considered resonances are expected to arise due to the coupling between the two epicyclic frequencies or between one epicyclic and the orbital frequency. The strongest resonance signals are excited for small integer ratios of the interacting frequencies, which reflects the properties of the observed QPOs and gives further support for the model.

The epicyclic frequencies for the Kerr black hole were first calculated in \cite{Aliev:1981,Aliev:1986} relating them to some resonant phenomena in the particle propagation in the equatorial plane. The topic was revisited later in \cite{Abramowicz:2001}-\cite{Torok:2005a} when resonance models were suggested as a possible explanation for the observed high frequency QPOs. Questioning the assumption that all the astrophysical black holes are described by the Kerr solution, QPOs were considered recently as one of the possible tools for probing the background spacetime. If we have independent measurements of the mass and the spin of some  astrophysical sources and if we assume that we are interpreting the data within the correct physical model, the observed values of the QPO frequencies can be used to differentiate between gravitational theories or between different types of compact objects. The possibility to test alternative  theories of gravity by means of the QPOs was studied in a number of recent works \cite{Stuchlik:2009}-\cite{Bambi:2012}. On the other hand broader classes of accreting compact objects were also considered \cite{Torok:2005b}-\cite{Deligianni:2021ecz} including naked singularities and traversable wormholes. The influence of the backreaction of the  accreting matter on the epicyclic motion was investigated in \cite{Nedkova:2020}  for the Schwarzschild black hole showing that it diversifies considerably the  resonance properties of the background spacetime.

In addition to black holes various further astrophysical candidates allow for accretion disks from which quasi-periodic oscillations might be observed. Wormholes attract a lot of attention specifically since they might mimic black holes in many respects, and numerous studies of their
potential astrophysical signatures have been performed in the recent years \cite{Damour:2007ap,Bambi:2013nla,Azreg-Ainou:2014dwa,Dzhunushaliev:2016ylj,Cardoso:2016rao, Konoplya:2016hmd,Nandi:2016uzg,Bueno:2017hyj,Blazquez-Salcedo:2018ipc}. Since long gravitational lensing by wormholes has been studied
\cite{Cramer:1994qj,Safonova:2001vz,Perlick:2003vg,Nandi:2006ds,Abe:2010ap,Toki:2011zu,Nakajima:2012pu,Tsukamoto:2012xs,Kuhfittig:2013hva,Bambi:2013nla,Takahashi:2013jqa,Tsukamoto:2016zdu},
the shadows of wormholes have been investigated
\cite{Bambi:2013nla,Nedkova:2013msa,Ohgami:2015nra,Shaikh:2018kfv,Gyulchev:2018fmd},
and, in particular, also accretion disks around wormholes have been explored
\cite{Harko:2008vy,Harko:2009xf,Bambi:2013jda,Zhou:2016koy,Lamy:2018zvj,Deligianni:2021ecz}.

The aim of this work is to study the properties of the epicyclic frequencies and the possible interpretation of the QPOs within the resonance models for some simple wormhole spacetimes by making comparison with the corresponding features for the Kerr black hole. In a recent work we considered in this respect some analytical geometries belonging to the Teo's class of traversable wormholes \cite{Deligianni:2021ecz}. We observed that some restrictions in the epicyclic motion which are typical for the Kerr black hole are not valid for these solutions. This enables the excitation of more diverse types of physically significant resonances. A natural question is whether a similar pattern will be repeated for another wormhole spacetime or whether it is possible  to isolate some general characteristic features in the resonance phenomena which can distinguish wormholes from black holes in respect to their QPOs.

For that purpose we consider some of the simplest wormholes within General Relativity obtained as solutions of the Einstein equations coupled to a phantom scalar field. The static solutions were obtained in analytical form by Ellis \cite{Ellis:1973yv} and Bronnikov \cite{Bronnikov:1973fh} and their properties were investigated in a number of works
\cite{Kodama:1978dw,Ellis:1979bh,Lobo:2005us,Lobo:2017oab}.
Their rotating generalizations have been constructed 
perturbatively \cite{Kashargin:2007mm,Kashargin:2008pk}
and numerically \cite{Kleihaus:2014dla,Chew:2016epf,Kleihaus:2017kai}.
Here we consider quasi-periodic oscillations in accretion disks around symmetric rotating Ellis wormholes
\cite{Kleihaus:2014dla,Chew:2016epf}.
We follow the discussion of the quasi-periodic oscillations considered 
before for a family of rotating Teo wormholes \cite{Teo:1998dp}
presented in \cite{Deligianni:2021ecz}, and compare with these results
as well as with the quasi-periodic oscillations in the accretion disks around the Kerr black hole.

The paper is organized as follows. We recall the properties of the rotating Ellis wormholes in section 2. We then derive the geodesic equations in section 3. We obtain the circular orbits,
and determine their domain of existence for the family of symmetric rotating Ellis wormholes.
In section 4 we derive the expressions for the epicyclic frequencies, and analyze the
stability of the circular orbits. We then address the ordering of
the orbital frequency, the radial epicyclic frequency and the vertical epicyclic frequency
in the domain of existence of circular orbits. Based on the different types of
ordering that are possible for these wormhole spacetimes we then consider in section 5
their parametric, forced and Keplerian resonances,
and compare with those of the Teo wormholes \cite{Deligianni:2021ecz}
and the Kerr black hole. In the last section we give our conclusions.

\section{Rotating Ellis Wormholes}

We briefly recall the rotating Ellis wormhole solutions of General Relativity 
\cite{Kleihaus:2014dla,Chew:2016epf}.
They are based on the action
\begin{eqnarray}
S= \frac{1}{16\pi G_{ }}\int d^4x \sqrt{-g} \left(R +
2g^{\mu\nu}\partial_{\mu}\psi \partial_{\nu}\psi 
  \right) \ ,
\end{eqnarray}
with phantom field $\psi$, and emerge 
as solutions of the coupled Einstein-scalar equations
\begin{eqnarray}
&&R_{\mu\nu} = - 2 \partial_\mu \psi \partial_\nu \psi \ , \\
&& \partial_\mu\left(\sqrt{-g}g^{\mu\nu}\partial_\nu \psi\right)  = 0 \ , 
\label{eqsi}
\end{eqnarray}
possessing two asymptotically flat regions connected by a throat.

The stationary rotating spacetimes are obtained with the line element
\begin{equation}
ds^2 = -e^{f} dt^2 + e^{-f} 
\left( e^{\nu} \left[d\eta^2 +h d\theta^2\right]
                    + h \sin^2\theta \left(d\phi -\omega dt\right)^2\right) \ ,
\label{lineel}
\end{equation}
where  $f$, $\nu$ and $\omega$ are functions of
the radial coordinate $\eta$, $-\infty< \eta < \infty$,
and the polar angle $\theta$, and $h=\eta^2 + \eta_0^2$ 
is an auxiliary function with parameter $\eta_0$.
Denoting the scalar charge by $D$,
the phantom field $\psi$ 
can be given in closed form
\begin{equation}
\psi = D \left(\arctan(\eta/\eta_0)-\pi/2\right) \ .
\label{exsol}
\end{equation}
The metric functions $f$, $\nu$ and $\omega$ are only known numerically
for rapidly rotating wormholes.
They satisfy the asymptotic boundary conditions
\begin{eqnarray}
&&\left. f\right|_{\eta \to \infty} = 0 \ , \ 
\left. \omega\right|_{\eta \to \infty} = 0 \ , \ 
\left. \nu\right|_{\eta \to \infty} = 0 \ , 
\label{bcinfty} \\
&&\left. f\right|_{\eta \to -\infty} = \gamma \ , \ 
\left. \omega\right|_{\eta \to -\infty} = \omega_{-\infty} \ , \ 
\left. \nu\right|_{\eta \to -\infty} = 0 \ .
\label{bcmininfty}
\end{eqnarray}
Symmetric wormholes satisfy $\gamma=0$.
For non-symmetric wormholes, on the other hand, $\gamma \ne 0$.
For static Ellis wormholes $\omega_{-\infty}=0$, and
\begin{equation}
f = \frac{\gamma}{2} \left(1-\frac{2}{\pi}\arctan\left(\frac{\eta}{\eta_0}\right) \right) \ , \
\omega = 0  \ , \ \nu = 0  \ . \
\label{statsol}
\end{equation}

We focus here on symmetric wormholes, where the metric functions
$f$ and $\nu$ are even functions of the radial coordinate $\eta$.
The wormhole throat resides at the center, described by the 
hypersurface $\eta=0=\eta_{th}$.

%
The equatorial circumferential radius $R_c(\eta)$ is defined as
\begin{equation}
R_c(\eta) = \left. \sqrt{g_{\varphi\varphi}}\right|_{\theta=\pi/2}
=\left. e^{-f/2} h \right|_{\theta=\pi/2} \ .
\label{Rcirc}
\end{equation}
The equatorial circumferential radius of the throat, which, for reasons of simplicity, we abbreviate as equatorial throat radius, is then given by 
$R_{e}=R_c(0)=\left. e^{-f/2} \eta_0 \right|_{\eta=0,\theta=\pi/2} $. 

We denote the angular velocity of the throat by
$\omega(\eta_{th})=\omega_{th}$ ,
and introduce the dimensionless rotational velocity
of the throat

\begin{equation}
v_e= R_e\omega_{th} \ .
\end{equation}

The mass $M$ and angular momentum $J$ of the wormholes
as measured in the $\eta > 0$ part of the spacetime
can be read off from the
components $g_{tt}$ and $g_{t\varphi}$,
\begin{equation}
g_{tt} \underset{\eta \to \infty} 
\longrightarrow - \left(1 - \frac{2 M}{\eta}\right)  \ , \ \ \
g_{t\varphi} \underset{\eta \to \infty} 
\longrightarrow -\frac{2 J  \sin^2\theta}{\eta} \ .
\label{mass_angmom}
\end{equation}
%
The symmetric wormholes satisfy a Smarr type relation
\cite{Kleihaus:2014dla,Chew:2016epf}
\begin{equation}
\label{smarr}
M =  2 \omega_{th} J \ .
\end{equation}
If $\omega_{th}$ is replaced by 
the horizon angular velocity $\omega_{H}$,
the relation agrees with the one for extremal Kerr black holes.

\begin{figure}[h!]
\begin{center}
\mbox{
\includegraphics[height=.24\textheight, angle =0]{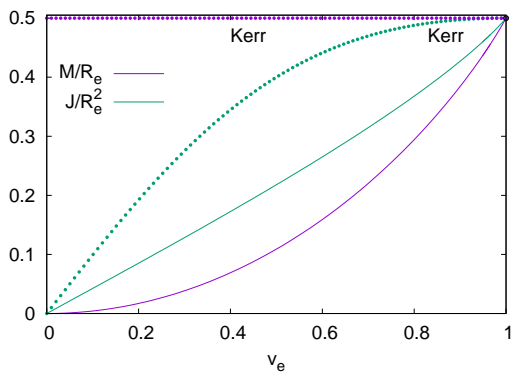}
}
\end{center}
\vspace{-0.5cm}
\caption{
Properties of symmetric rotating wormholes at fixed equatorial throat radius $R_e$:
Scaled mass $M/R_e$ and scaled angular momentum $J/R_e^2$
vs rotational velocity $v_e$ of the throat in the equatorial plane.
For comparison the analogous quantities ($R_e \to R_H$) are shown
for Kerr black holes.
The black dot highlights the extremal Kerr values.}
\label{fig1}
\end{figure}

In Fig.~\ref{fig1} we show the scaled mass $M/R_e$ and scaled angular momentum $J/R_e^2$
versus the rotational velocity of the throat in the equatorial plane, $v_e$.
Both quantities rise monotonically with increasing $v_e$ from the static case $v_e=0$
to the maximal rotational velocity $v_e=1$.
For comparison, we also show the global charges for
Kerr black holes, now scaled with the horizon radius in the equatorial plane, $R_H$.
The figure clearly demonstrates the convergence of the wormhole values
to the extremal Kerr values, when $v_e \to 1$.
In fact, as studied in detail in \cite{Kleihaus:2014dla,Chew:2016epf},
the full wormhole solutions converge to the extremal Kerr solutions in the limit,
not only the global charges.

\section{Analysis of the circular orbits}

We briefly recall the general formulae for some kinematic quantities of circular orbits in 
a stationary, axisymmetric spacetime
\begin{eqnarray}
ds^2=g_{tt}\,dt^2+2g_{t\phi}\,dt d\phi+g_{\eta\eta}\,d\eta^2+g_{\theta\theta}\,d\theta^2+g_{\phi\phi}\,d\phi^2\,.
\end{eqnarray}
For motion along geodesics the constraint $g_{\mu\nu}{\dot x}^{\mu}{\dot x}^{\nu} =\epsilon$ holds,
where $\epsilon=-1$ ($\epsilon = 0$) for massive (massless) particles.
Because of the spacetime symmetries, the specific energy $E$ and angular momentum $L$ of the particles are conserved.
The geodesic equations in the equatorial plane for massive particles are given by
\begin{eqnarray}
\frac{dt}{d\tau}&=&\frac{E g_{\phi\phi}+L g_{t\phi}}{g_{t\phi}^2-g_{tt}g_{\phi\phi}} \ ,   \nonumber   \\[1mm]
\frac{d\phi}{d\tau}&=&-\frac{E g_{t\phi}+ L g_{tt}}{g_{t\phi}^2-g_{tt}g_{\phi\phi}} \ , \nonumber \\[1mm]
    g_{\eta\eta}\left(\frac{d\eta}{d\tau}\right)^2&=&-1+\frac{E^2 g_{\phi\phi}+2 E L g_{t\phi}+ L^2g_{tt}}{g_{t\phi}^2-g_{tt}g_{\phi\phi}} 
= V_{\rm eff} \ ,
    \label{geodeqs3}
\end{eqnarray}
where we have introduced an affine parameter $\tau$ and the effective potential $V_{\rm eff}$.
Circular orbits correspond to the stationary points of
 $V_{\rm eff}$
 
\begin{eqnarray}
V_{\rm eff}(\eta)=0 \ , \quad~~~ V_{\rm eff,\eta}(\eta)=0 \ ,
\end{eqnarray}
where we have denoted the derivative with respect to a coordinate by a comma.
For the circular orbits these conditions lead to the specific energy and angular momentum 
\begin{eqnarray}
E &=&-\frac{g_{tt}+g_{t\phi}\omega_0}{\sqrt{-g_{tt}-2g_{t\phi}\omega_0-g_{\phi\phi}\omega_0^2}} \ ,   
 \nonumber  \\[2mm]
L &=& \frac{g_{t\phi}+g_{\phi\phi}\omega_0}{\sqrt{-g_{tt}-2g_{t\phi}\omega_0-g_{\phi\phi}\omega_0^2}} \ ,     \nonumber
\end{eqnarray}
and the angular velocity 
\begin{eqnarray}
\omega_0&=&\frac{d\phi}{dt}=\frac{-g_{t\phi,\eta}\pm \sqrt{(g_{t\phi,\eta})^2
                            -g_{tt,\eta}g_{\phi\phi,\eta}}}{g_{\phi\phi,\eta}},
\label{Omega}
\end{eqnarray}
where the $+$ sign ($-$ sign) refers to the co-rotating (counter-rotating) orbits.
In terms of the metric parametrization (\ref{lineel}) these expressions become
\begin{eqnarray}
E &=& \frac{e^f + e^{-f}h\omega(\omega_0-\omega) }{\sqrt{e^f - e^{-f}h(\omega_0 -\omega)^2}}\ , 
\nonumber \\[2mm]
L &=& \frac{e^{-f}h(\omega_0 -\omega)}{\sqrt{e^f - e^{-f}h(\omega_0 -\omega)^2}} \ , 
\nonumber \\[2mm] 
\omega_0 &=& \omega +\frac{e^{-f}h\omega_{,\eta} 
   \pm \sqrt{e^f_{\ ,\eta}(e^{-f}h)_{,\eta} + (e^{-f}h)^2(\omega_{,\eta})^2}}{(e^{-f}h)_{,\eta}} \ . 
   \label{omega_0} 
\end{eqnarray}
In order for timelike circular orbits to exist the inequality $e^f - e^{-f}h(\omega_0 -\omega)^2>0$ 
should be satisfied.

\begin{figure}[t!]
\begin{center}
(a)\mbox{
\includegraphics[width=.45\textwidth, angle =0]{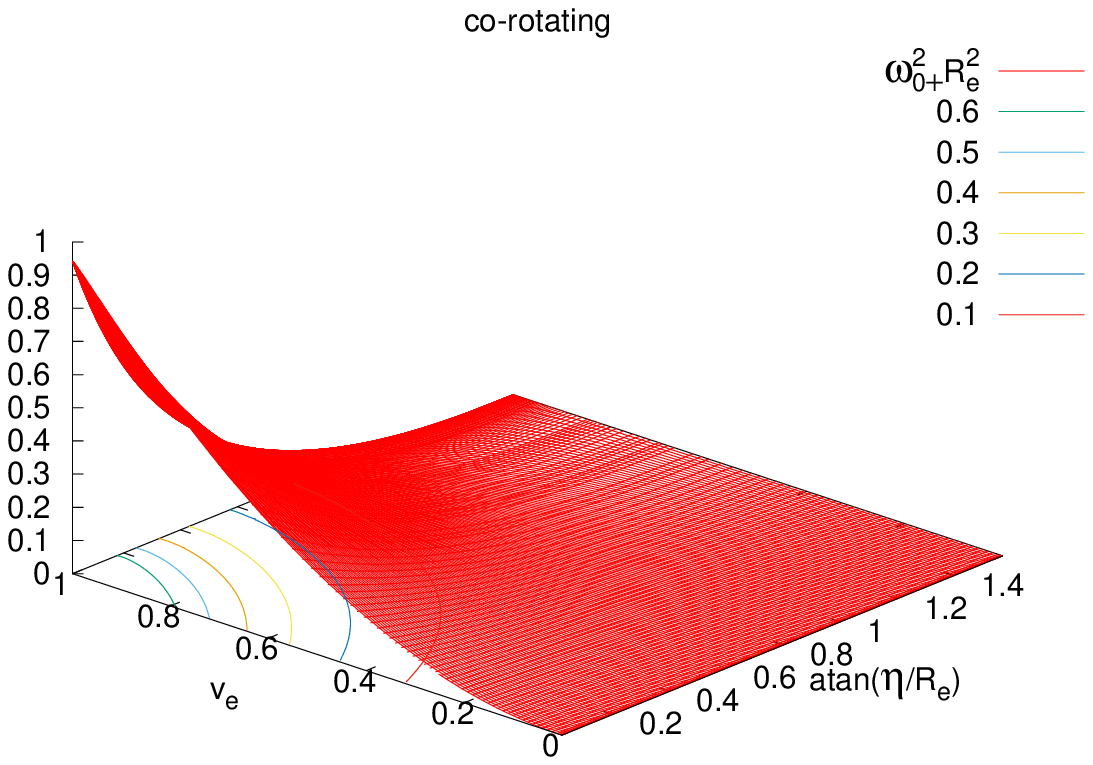} 
(b)      
\includegraphics[width=.45\textwidth, angle =0]{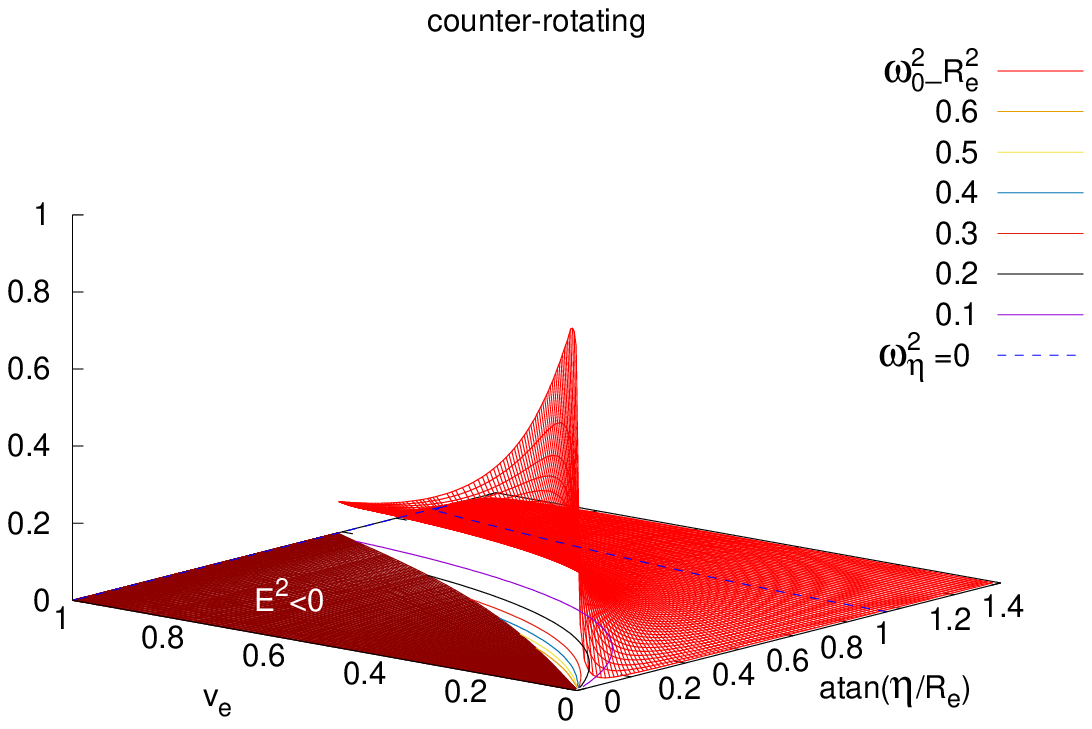}
}
\end{center}
\caption{
Scaled squared angular velocities at fixed equatorial throat radius $R_e$
for co-rotating circular orbits, $\omega^2_{0+} R_e^2$ (a),
and for counter-rotating circular orbits, $\omega^2_{0-} R_e^2$ (b),
vs throat velocity $v_e$ and compactified scaled radial coordinate 
$\atan (\eta/R_e) \ge 0$.
The curves in the $v_e-\atan (\eta/R_e)$ plane represent a set of contour lines.
\label{fig2}
}
\end{figure}

We now turn to the numerical calculations for the circular orbits.
Since the specific angular momentum $a=J/M$ diverges for symmetric wormholes in
the static limit, we do not consider the domain of existence of circular orbits 
in terms of $a/M$.
Instead we employ the throat velocity in the equatorial plane $v_e$.
We recall that for symmetric wormholes the throat velocity $v_e$ ranges from
$v_e=0$ in the static case to $v_e \to 1$ in the critical case,
where in the limit the extremal Kerr black hole is approached.

We exhibit the squared angular velocities $\omega_0^2$ in Fig.\ref{fig2}.
To obtain dimensionless quantities, we scale the angular velocities
and also the radial coordinate $\eta$ by the equatorial throat radius $R_e$.
Fig.\ref{fig2}(a) then shows the scaled squared angular velocities
$\omega^2_{0+} R_e^2$ for co-rotating circular orbits
versus the throat velocity $v_e$ and the compactified scaled radial coordinate 
$\atan (\eta/R_e) \ge 0$.
Fig.\ref{fig2}(b) is the analogous plot for the 
counter-rotating circular orbits, showing $\omega^2_{0-} R_e^2$.
The three-dimensional figures also contain a set of contour lines.

\begin{figure}[t!]
\begin{center}
\mbox{\includegraphics[width=.47\textwidth, angle =0]{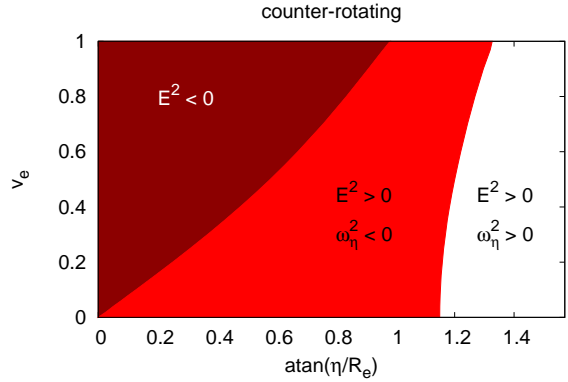}
}
\end{center}
\caption{Domain of existence of counter-rotating circular orbits
in the $v_e-\atan (\eta/R_e)$ plane:
region with stable (white), unstable (light-red) and no (dark-red) counter-rotating circular orbits.
\label{fig3}
}
\end{figure}

We observe that co-rotating circular orbits exist in the full domain
$0<v_e < 1$, $ 0 < \eta/R_e < \infty$.
An analysis of the limiting cases $v_e=0$ (static wormholes)
and $\eta=0$ (throat) is performed in the Appendix.
The situation differs for the counter-rotating circular orbits.
They exist only in a part of the domain, as illustrated in Fig.\ref{fig3}.
Thus, there are no counter-rotating circular orbits
in the dark-red part of the  $v_e- \atan  (\eta/R_e)$ plane,
labeled by unphysical energies, $E^2< 0$.

\section{Epicyclic frequencies}

To derive the epicyclic frequencies we start from the geodesic equations in the equatorial plane
\begin{equation}\label{geodesics}
\ddot{x}^{\alpha}+\Gamma^{\alpha}_{\,\,\,\beta\gamma}\dot{x}^{\beta}\dot{x}^{\gamma}=0\,,
\end{equation}
and slightly perturb the circular motion $\tilde{x}^{\mu}(s) = x^{\mu}(s) + \xi^\mu(s)$, where $x^{\mu}(s)$  denotes the circular orbit and $s$ is the affine parameter. In the linear approximation one then finds  \cite{Aliev:1980hz,Aliev:1986wu}
\begin{eqnarray}\label{pert}
&&\frac{d^2\xi^\mu}{dt^2} + 2\gamma^\mu_\alpha\frac{d\xi^\alpha}{dt} + \xi^b\partial_b{\cal V}^{\mu} = 0\, , \quad b = \eta, \theta \nonumber \\[2mm]
&&\gamma^\mu_\alpha =\left[\Gamma^\mu_{\alpha\beta} u^\beta(u^0)^{-1}\right]_{\theta=\pi/2}\, , \nonumber \\[2mm]
&& {\cal V}^{\mu} = \left[\gamma^\mu_\alpha u^\alpha(u^0)^{-1}\right]_{\theta=\pi/2},
\end{eqnarray}
with 4-velocity $u^\mu = \dot{x^\mu}= u^0(1, 0, 0, \omega_0)$. 
Integrating the equations for the $t$ and $\phi$ perturbations leads to 
\begin{eqnarray}
\frac{d\xi^A}{dt} + 2\gamma^A_\alpha\xi^\alpha = 0\, , \quad A = t,\phi .
\end{eqnarray}
The equations for the radial and vertical perturbations then decouple and read
\begin{eqnarray}\label{freq}
&&\frac{d^2\xi^\eta}{dt^2} + \omega_\eta^2\xi^\eta = 0\, , \label{pertx} \\[2mm]
&&\frac{d^2\xi^\theta}{dt^2} + \omega_\theta^2\xi^\theta = 0\, ,  \nonumber
\end{eqnarray}
with epicyclic frequencies $\omega_\eta$ and $\omega_\theta$,
\begin{eqnarray}
&&\omega_\eta^2 = \partial_\eta {\cal V}^\eta - 4\gamma^\eta_A\gamma^A_\eta, \\[2mm] \nonumber
&&\omega_\theta^2 = \partial_\theta{\cal V}^\theta. \nonumber
\end{eqnarray}

Evaluating these expressions for the epicyclic frequencies yields

\begin{eqnarray}
\omega^2_\eta  & = &
 \frac{1}{2 h  e^{2f +\nu}}
\left[ 2e^{2f} \omega_0^2 (4 \eta^2-h)-2 h^3 (\omega_0-\omega)^2 \omega_{,\eta}^2
  -e^{4 f} h (f_{,\eta} \nu_{,\eta}-f_{,\eta\eta})
\right.
\nonumber\\ & &
\left.  
  +e^{2f}\left\{\omega^2 (8 \eta^2-2 h+
       h (2 h f_{,\eta}^2+2\eta \nu_{,\eta}-f_{,\eta}(6\eta+h\nu_{,\eta})+h f_{,\eta\eta}))
\right.
\right.
\nonumber\\ & &
\left.  
\left.  
   -2\omega \left(\omega_0 (8 \eta^2-2h)+h \left(2 \omega_0 h f_{,\eta}^2
   +f_{,\eta}(-6 \omega_0\eta-h (\omega_0 \nu_{,\eta}+\omega_{,\eta}))
\right.
\right.
\right.
\right.
\nonumber\\ & &
\left.  
\left.  
\left.  
\left.  
   +\nu_{,\eta} (2 \omega_0 \eta-h \omega_{,\eta})+h (\omega_0 f_{,\eta\eta}+\omega_{,\eta\eta})
   \right)\right)
\right.
\right.
\nonumber\\ & &
\left.  
\left.  
   +\omega_0 h (2 \omega_0 h f_{,\eta}^2+f_{,\eta} (-6 \omega_0 \eta-h (\omega_0 \nu_{,\eta}+2\omega_{,\eta}))
   +2 \nu_{,\eta} (\omega_0 \eta-h \omega_{,\eta})+h (\omega_0 f_{,\eta\eta}+2\omega_{,\eta\eta}))\right\}\right]
\end{eqnarray}
\begin{eqnarray}
\omega^2_\theta & = & \frac{1}{2 h e^\nu}(e^{2f} f_{,\theta\theta}
           +h (\omega_0-\omega)((\omega_0-\omega) (2+f_{,\theta\theta})+2\omega_{,\theta\theta}))
\end{eqnarray}
Alternative expressions for the frequencies are presented in the Appendix.

\begin{figure}[t!]
\begin{center}
(a)\mbox{
\includegraphics[width=.45\textwidth, angle =0]{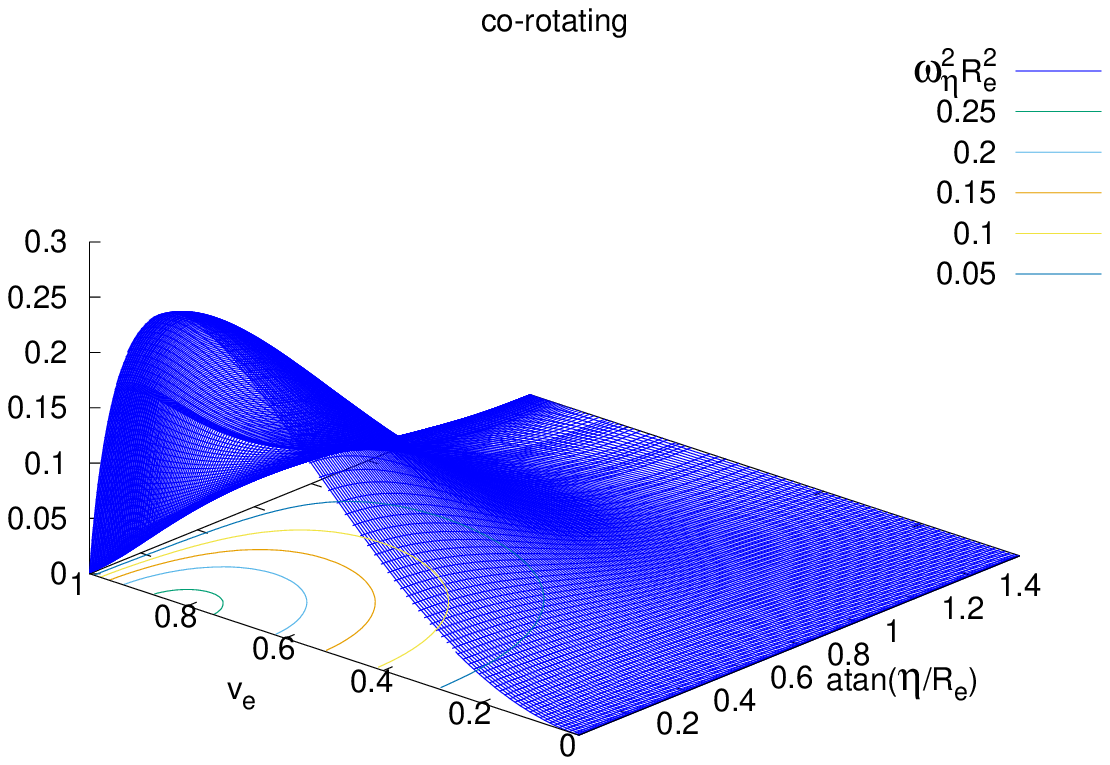} 
(b)      
\includegraphics[width=.45\textwidth, angle =0]{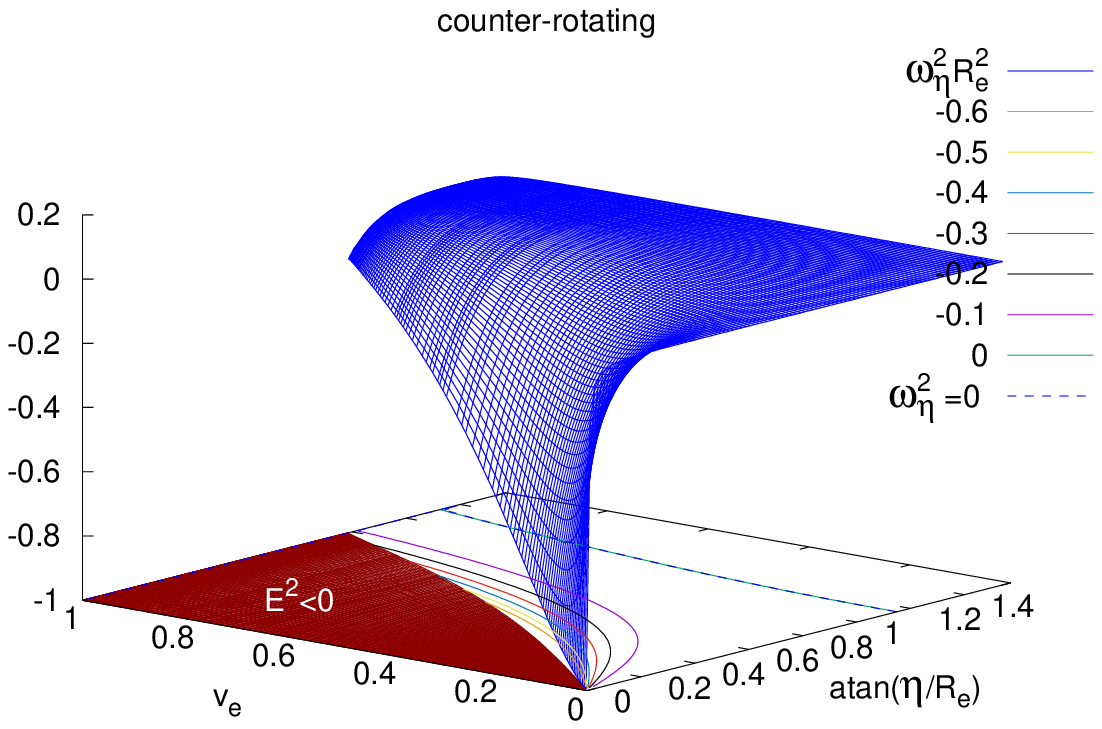}
}
\end{center}
\caption{
Scaled squared radial epicyclic frequencies $\omega^2_\eta R_e^2$ 
at fixed equatorial throat radius $R_e$
for co-rotating circular orbits (a),
and counter-rotating circular orbits (b),
vs throat velocity $v_e$ and compactified scaled radial coordinate 
$\atan (\eta/R_e) \ge 0$.
The curves in the $v_e-\atan (\eta/R_e)$ plane represent a set of contour lines.
\label{fig4}
}
\end{figure}
\begin{figure}[t!]
\begin{center}
(a)\mbox{
\includegraphics[width=.45\textwidth, angle =0]{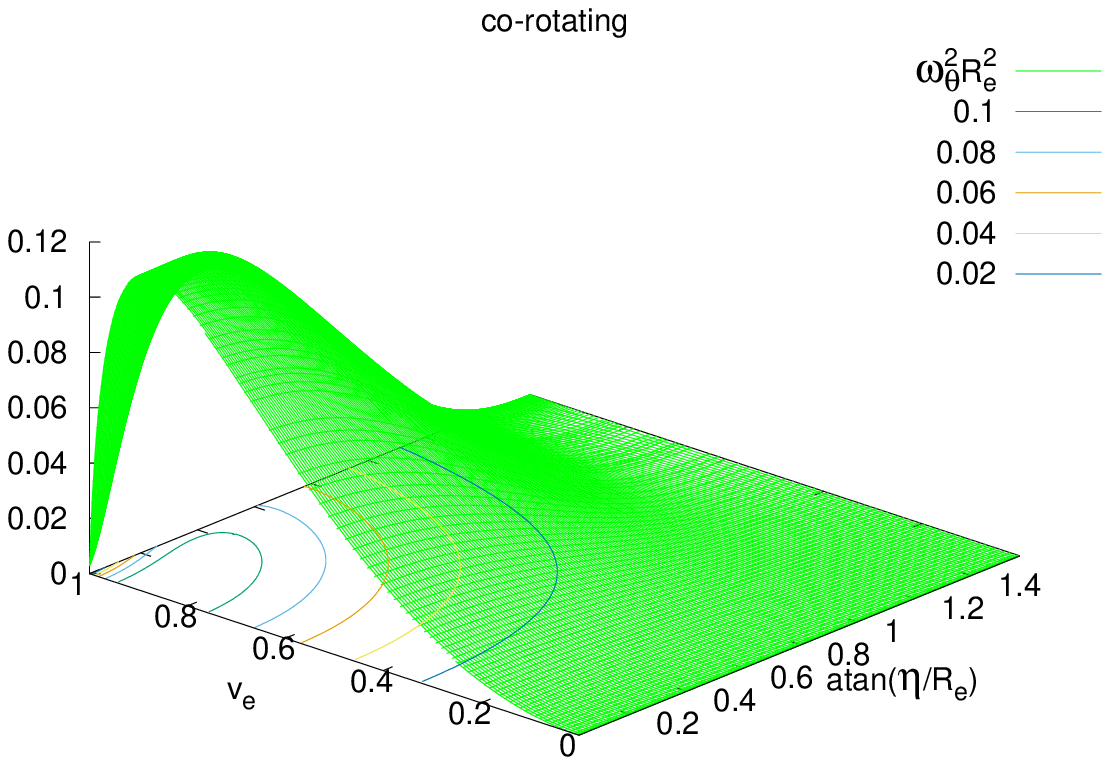} 
(b)      
\includegraphics[width=.45\textwidth, angle =0]{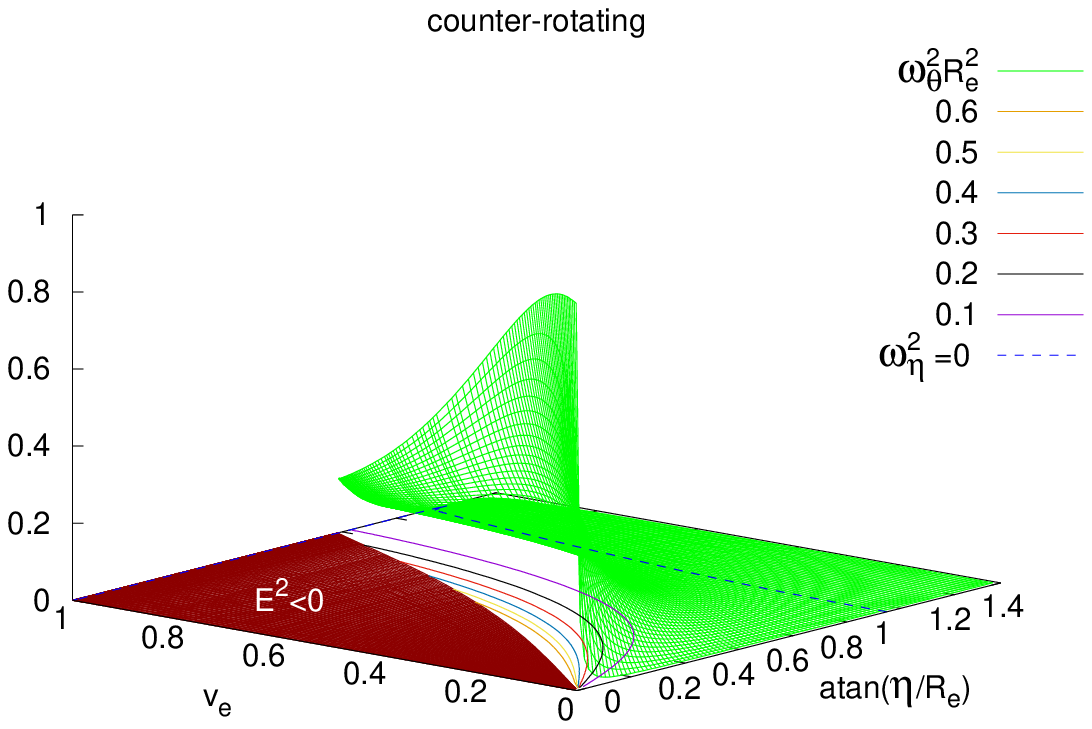}
}
\end{center}
\caption{
Scaled squared vertical epicyclic frequencies $\omega^2_\theta R_e^2$ 
at fixed equatorial throat radius $R_e$
for co-rotating circular orbits (a),
and counter-rotating circular orbits (b),
vs throat velocity $v_e$ and compactified scaled radial coordinate 
$\atan (\eta/R_e) \ge 0$.
The curves in the $v_e-\atan (\eta/R_e)$ plane represent a set of contour lines.
\label{fig5}
}
\end{figure}

We exhibit the squared radial epicyclic frequencies $\omega_\eta^2$ in Fig.\ref{fig4}.
Again, we have scaled the frequencies with the equatorial throat radius $R_e$
to obtain dimensionless quantities.
Fig.\ref{fig4}(a) exhibits the radial epicyclic frequencies for co-rotating orbits,
and Fig.\ref{fig4}(b) for counter-rotating orbits
versus the throat velocity $v_e$ and the compactified scaled radial coordinate 
$\atan (\eta/R_e) \ge 0$.
The figures also show a set of contour lines.

For the co-rotating circular orbits 
the squared radial epicyclic frequencies $\omega_\eta^2$
are positive in the domain
$0<v_e < 1$, $ 0 < \eta/R_e < \infty$.
In the limiting cases $v_e=0$ (static wormholes)
and $\eta=0$ (throat) zero modes may arise as discussed in the Appendix.
For the counter-rotating circular orbits, however,
the squared radial epicyclic frequencies $\omega_\eta^2$
are positive only in a part of their domain of existence.
The zero modes are indicated by a contour line in Fig.\ref{fig4}(b).
The domain of unstable counter-rotating circular orbits is highlighted 
in light-red in Fig.\ref{fig3}.

The squared vertical epicyclic frequencies $\omega_\theta^2$ are shown in Fig.\ref{fig5}.
They are positive for the co-rotating circular orbits in the domain
$0<v_e < 1$, $ 0 < \eta/R_e < \infty$, as seen in  Fig.\ref{fig5}(a).
Fig.\ref{fig5}(b) demonstrates, that within the domain of existence
of counter-rotating orbits inside the region $0<v_e < 1$, $ 0 < \eta/R_e < \infty$,
also the counter-rotating circular orbits possess positive 
squared vertical epicyclic frequencies $\omega_\theta^2$.
Thus the vertical epicyclic frequencies do not lead to further instabilities
of the circular orbits.
Again, in the limiting cases $v_e=0$ (static wormholes)
and $\eta=0$ (throat) zero modes may arise (see Appendix).

\begin{figure}[h!]
\begin{center}
\mbox{
\includegraphics[height=.24\textheight, angle =0]{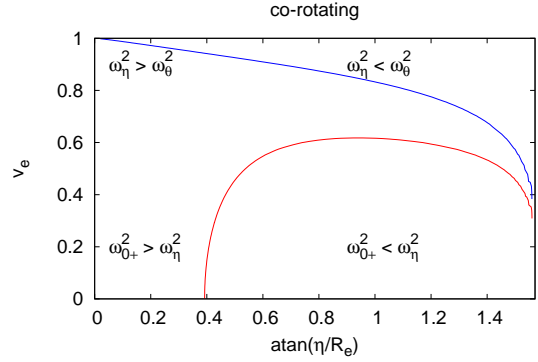}
}
\end{center}
\vspace{-0.5cm}
\caption{
Ordering of the frequencies in the $v_e-\atan (\eta/R_e)$ plane
for co-rotating orbits:
equality of the orbital frequency $\omega_0$ and the
radial epicyclic frequency $\omega_\eta$ (red curve),
and
equality of the radial epicyclic frequency $\omega_\eta$
and the  vertical epicyclic frequency $\omega_\theta$ (blue curve).
}
\label{fig6}
\end{figure}

We now consider the ordering of the frequencies.
We demonstrate the equality of frequencies for co-rotating orbits in the 
$v_e-\atan (\eta/R_e)$ plane in Fig.\ref{fig6}.
The red curve indicates 
equality of the orbital frequency $\omega_{0+}$ and the
radial epicyclic frequency $\omega_\eta$,
while the blue curve shows the
equality of the radial epicyclic frequency $\omega_\eta$
and the vertical epicyclic frequency $\omega_\theta$
for co-rotating orbits.
These curves divide the domain of existence into three regions.
In the region below the red curve we observe the ordering
$\omega^2_\eta \geq \omega^2_{0+} \geq \omega^2_\theta$.
Between the red and blue curves the ordering changes to
$\omega^2_{0+} \geq \omega^2_\eta \geq  \omega^2_\theta$,
and above the blue curve the ordering is
$\omega^2_{0+}  \geq  \omega^2_\theta \geq \omega^2_\eta$.

The latter ordering corresponds to the ordering found for the Kerr
black holes. This is expected, since the wormhole spacetimes are approaching the
extremal Kerr spacetime in this region.
The family of rotating Teo wormholes discussed in \cite{Deligianni:2021ecz}
exhibits also all three types of orderings. 
However, they occur in different regions of the domain of existence
as compared to the symmetric rotating Ellis wormholes.
For counter-rotating orbits the ordering is always
$\omega^2_\theta \geq \omega^2_{0-} \geq \omega^2_\eta$.
This holds for the symmetric rotating Ellis wormholes,
the family of rotating Teo wormholes \cite{Deligianni:2021ecz},
and the Kerr black holes.

\begin{figure}[h!]
\begin{center}
\mbox{
\includegraphics[height=.24\textheight, angle =0]{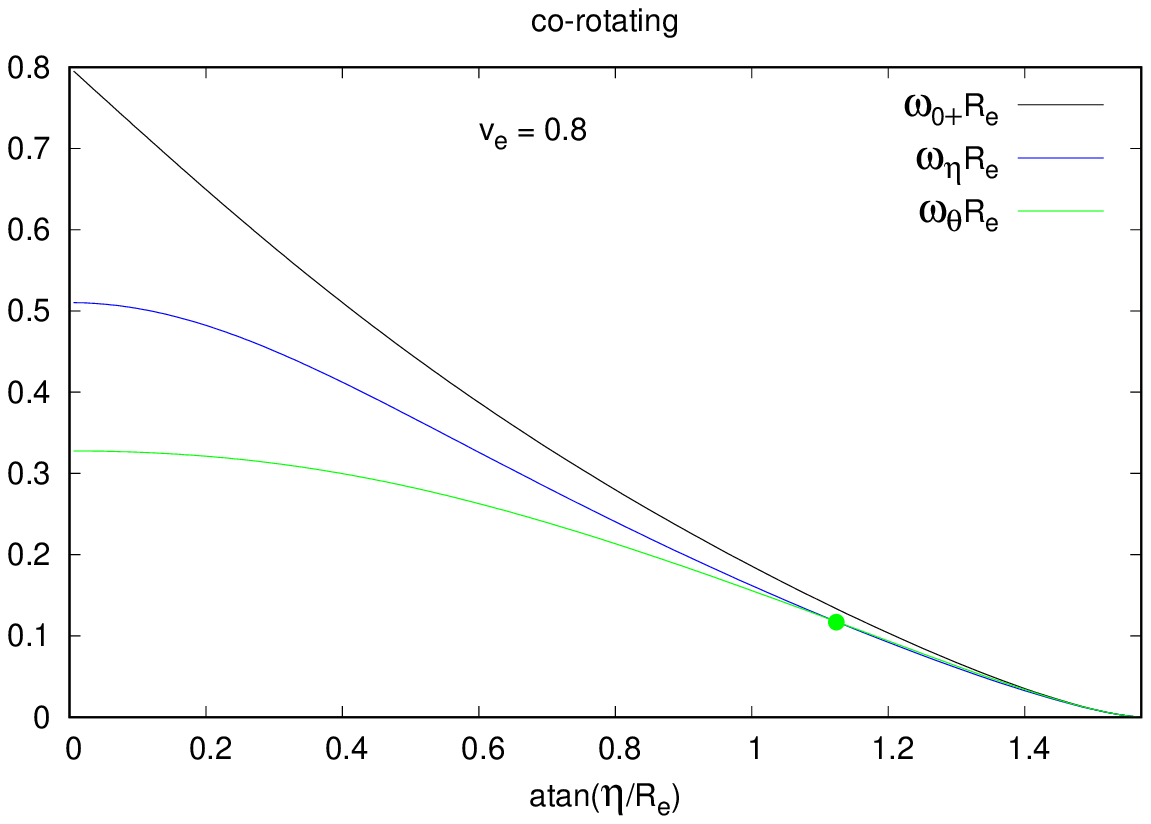}
\includegraphics[height=.24\textheight, angle =0]{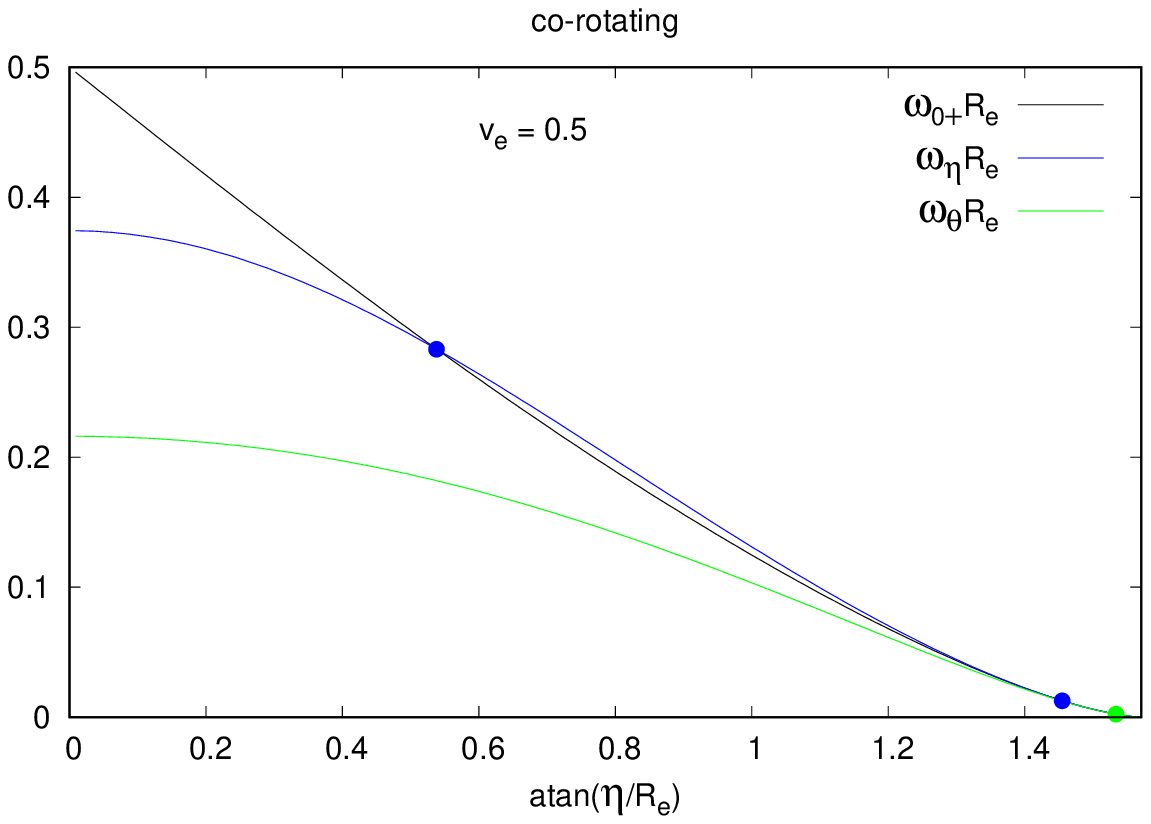}
}
\mbox{
\includegraphics[height=.24\textheight, angle =0]{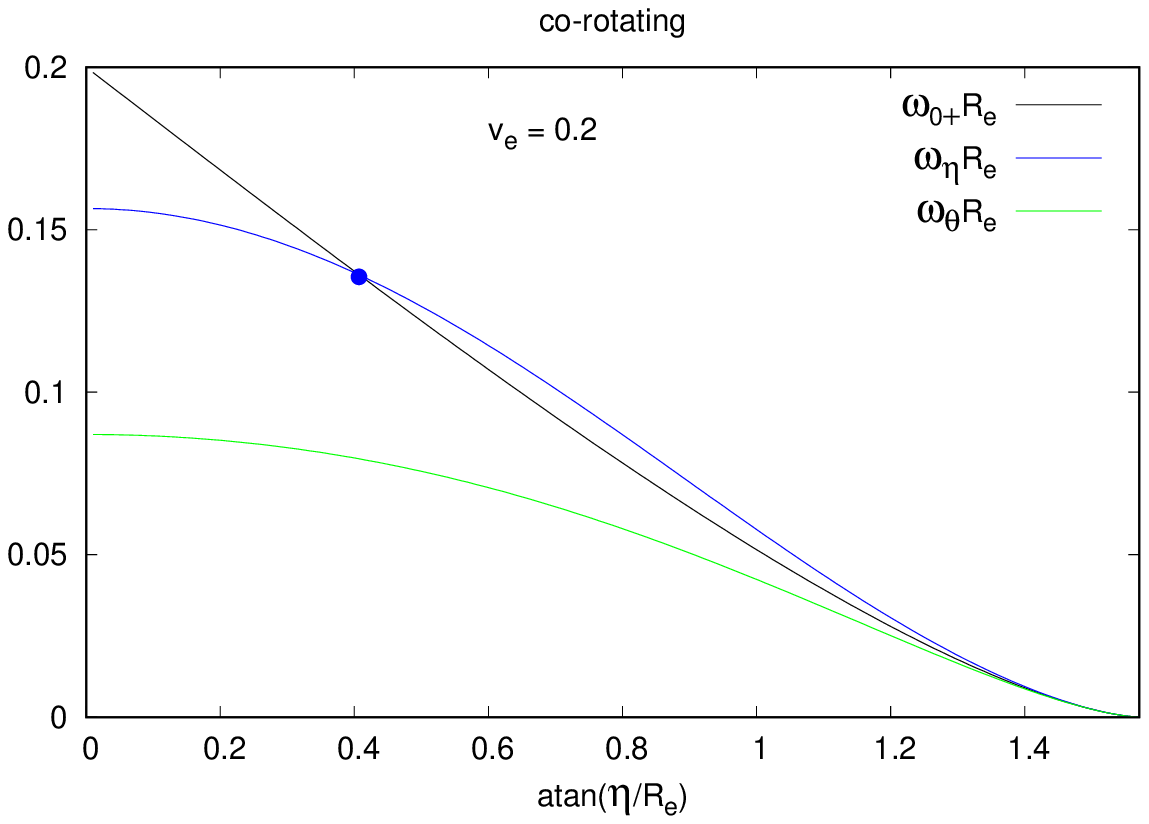}
\includegraphics[height=.24\textheight, angle =0]{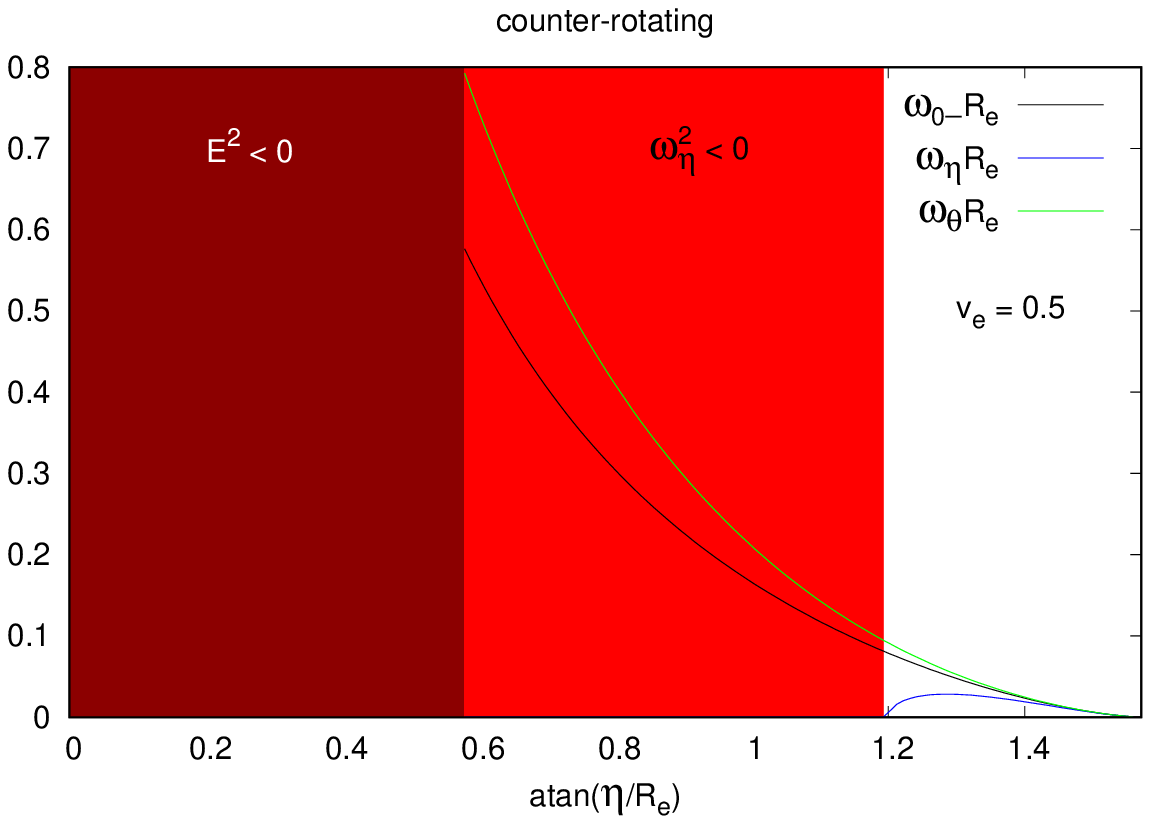}
}
\end{center}
\vspace{-0.5cm}
\caption{
Ordering of the frequencies:
scaled orbital frequencies $\omega_0 R_e$,
radial epicyclic frequencies $\omega_\eta R_e$,
and vertical epicyclic frequencies $\omega_\theta R_e$
vs compactified scaled radial coordinate 
$\atan (\eta/R_e)$
for wormholes with $v_e = 0.8$ (a), $v_e = 0.5$ (b), $v_e = 0.2$ (c)
for co-rotating orbits,
and $v_e = 0.5$ (d) for counter-rotating orbits.
The dots indicate orbits with equal orbital and radial epicyclic frequencies
(blue), respectively equal vertical and radial epicyclic frequencies (green).
}
\label{fig7}
\end{figure}

The various orderings of the frequencies can also be seen in Fig.\ref{fig7},
where the scaled orbital frequencies $\omega_0 R_e$,
radial epicyclic frequencies $\omega_\eta R_e$,
and vertical epicyclic frequencies $\omega_\theta R_e$
are exhibited as functions of the compactified scaled radial coordinate 
$\atan (\eta/R_e)$ for several wormhole solutions.
Fig.\ref{fig7}(a)-(c) show the frequencies for co-rotating orbits
for wormholes with $v_e = 0.8$ (a), $v_e = 0.5$ (b), and $v_e = 0.2$ (c),
while Fig.\ref{fig7}(d) shows the frequencies for counter-rotating orbits
for a wormhole with $v_e = 0.5$ (d). 
The dots in the figures
highlight the changes in the ordering, when a line of frequency equality is crossed.

\begin{figure}[h!]
\begin{center}
\mbox{
\includegraphics[height=.24\textheight, angle =0]{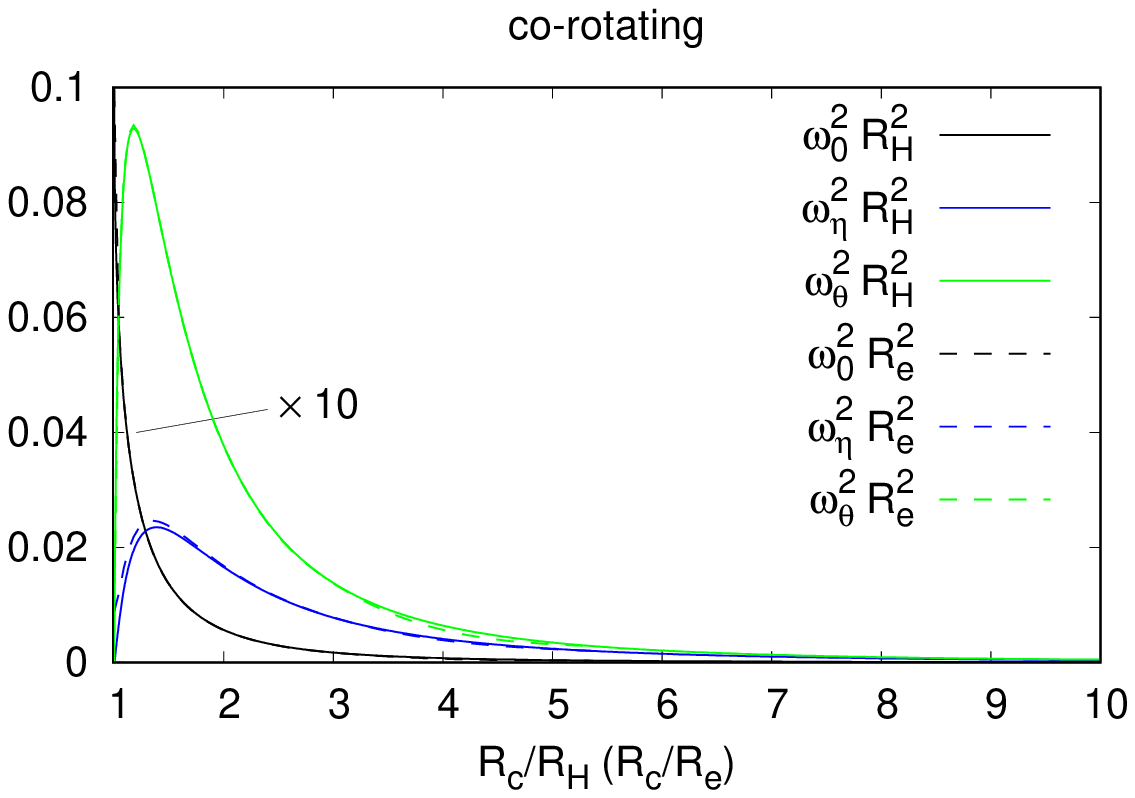}
\includegraphics[height=.24\textheight, angle =0]{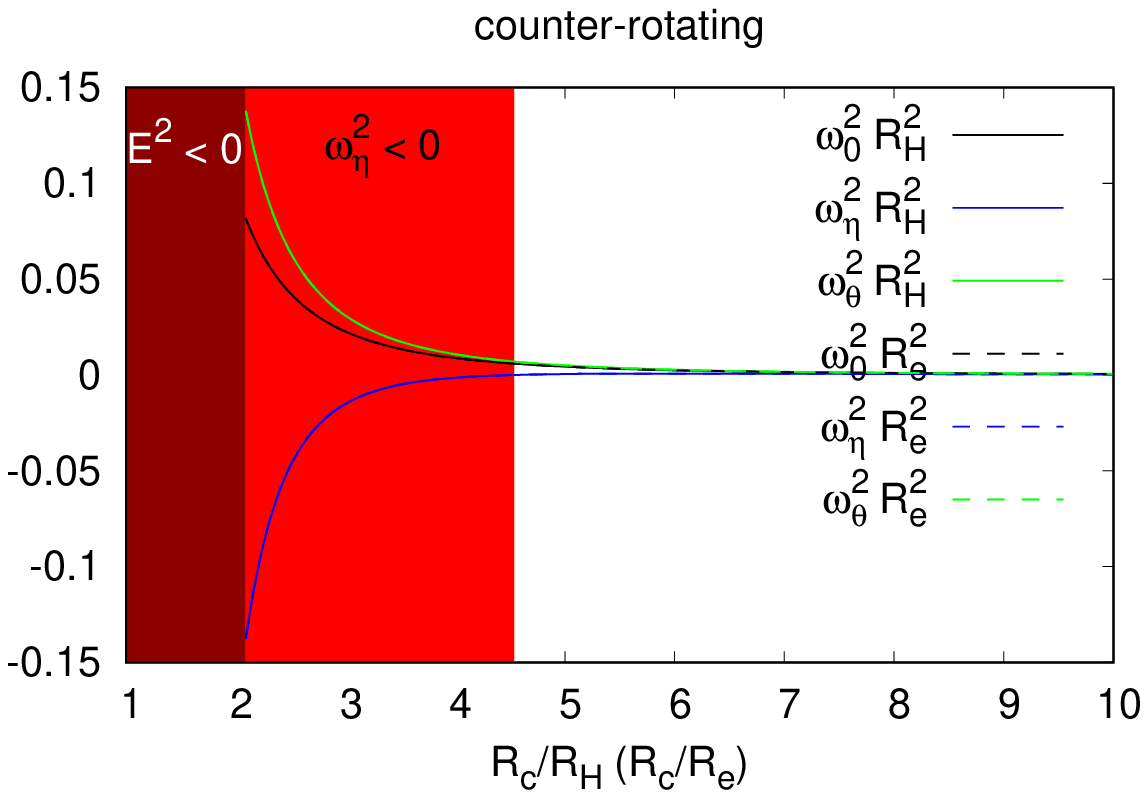}
}
\end{center}
\vspace{-0.5cm}
\caption{
Rapidly rotating limit:
scaled squared orbital frequencies $\omega_0^2 R_e^2$
($\omega_0^2 R_H^2$),
radial epicyclic frequencies $\omega_\eta^2 R_e^2$
($\omega_\eta^2 R_H^2$),
and vertical epicyclic frequencies $\omega_\theta^2 R_e^2$
($\omega_\eta^2 R_H^2$),
vs scaled circumferential radius $R_c/R_e$
($R_c/R_H$)
for a wormhole with $v_e = 0.9988$
(for an extremal Kerr black hole with horizon radius $R_H$)
for co-rotating orbits (a) and counter-rotating orbits (b).
}
\label{fig8}
\end{figure}

We demonstrate in Fig.\ref{fig8}, that the orbital frequencies and the
epicyclic frequencies converge to the respective frequencies of the
extremal Kerr black hole, as could have been expected from the previous studies of 
the rapidly rotating limit of the wormholes.
Here we have chosen a velocity $v_e = 0.9988$ of the wormhole throat
that is very close to the limiting value $v_e=1$,
and we have replaced the circumferential radius $R_e$ of the throat
in  the equatorial plane by the respective circumferential radius $R_H$
of the black  hole horizon.

\section{Analysis of 
Quasi-periodic Oscillations}

We now recall two simple generic models to generate resonances.
To that end we consider non-linear corrections to the perturbation equations 
for the geodesic circular motion
\begin{eqnarray}
&&\frac{d^2 \xi^\eta}{dt^2} + \omega_\eta^2 \xi^\eta =  \omega_\eta^2 f_\eta \left(\xi^\eta, \xi^\theta, \frac{d\xi^\eta}{dt}, \frac{d\xi^\theta}{dt}\right), \nonumber \\[2mm]
&&\frac{d^2 \xi^\theta}{dt^2} + \omega_\theta^2 \xi^\theta =  \omega_\theta^2 f_\theta\left(\xi^\eta, \xi^\theta, \frac{d\xi^\eta}{dt}, \frac{d\xi^\theta}{dt}\right) ,
\end{eqnarray}
where $f_\eta$ and $f_\theta$ are non-linear functions. 
(i) We assume, that $f_\eta=0$ and  $f_\theta = h\,\xi^\eta \xi^\theta$, with coupling constant $h$. 
This yields for the vertical oscillations
\begin{eqnarray}
\frac{d^2 \xi^\theta}{dt^2} + \omega_\theta^2 \xi^\theta = - \omega_\theta^2 h \cos(\omega_\eta t) \xi^\theta \ ,
\end{eqnarray}
and thus the Mathieu equation, which describes \textit{parametric resonances},
where the frequencies possess ratio
\begin{eqnarray}
\frac{\omega_\eta}{\omega_\theta}= \frac{2}{n} \ ,
\end{eqnarray}
with positive integer $n$ (see e.g. \cite{Landau}). 
As shown in \cite{Abramowicz:2003,Rebusco,Horak:2004a,Horak:2004b}, parametric resonances
represent a mathematical property of thin, nearly Keplerian disks.
(ii) We consider a forced non-linear oscillator, assuming a periodic radial force for the vertical oscillations
with frequency equal to the radial epicyclic frequency
\begin{eqnarray}
\frac{d^2 \xi^\theta}{dt^2} + \omega_\theta^2 \xi^\theta
+ [{\rm non \; linear \; terms \; in} \: \xi^\theta] &=&
h(\eta) \cos(\omega_\eta t) \ .
\end{eqnarray}
This leads to \textit{forced resonances} with a frequency ratio
\begin{eqnarray}
\frac{\omega_\eta}{\omega_\theta}= \frac{m}{n} \ ,
\end{eqnarray}
where $m$ and $n$ are positive integers, but also linear combinations of the epicyclic frequencies may be considered,
yielding further options to fit the quasiperiodic oscillations.
Besides these resonances due to the coupling of the epicyclic frequencies,
it is also possible, though physically less motivated,
to consider \textit{Keplerian resonances} that might occur due to the interaction
of an epicyclic frequency with the orbital frequency.


\begin{figure}[h!]
\begin{center}
\mbox{
\includegraphics[height=.24\textheight, angle =0]{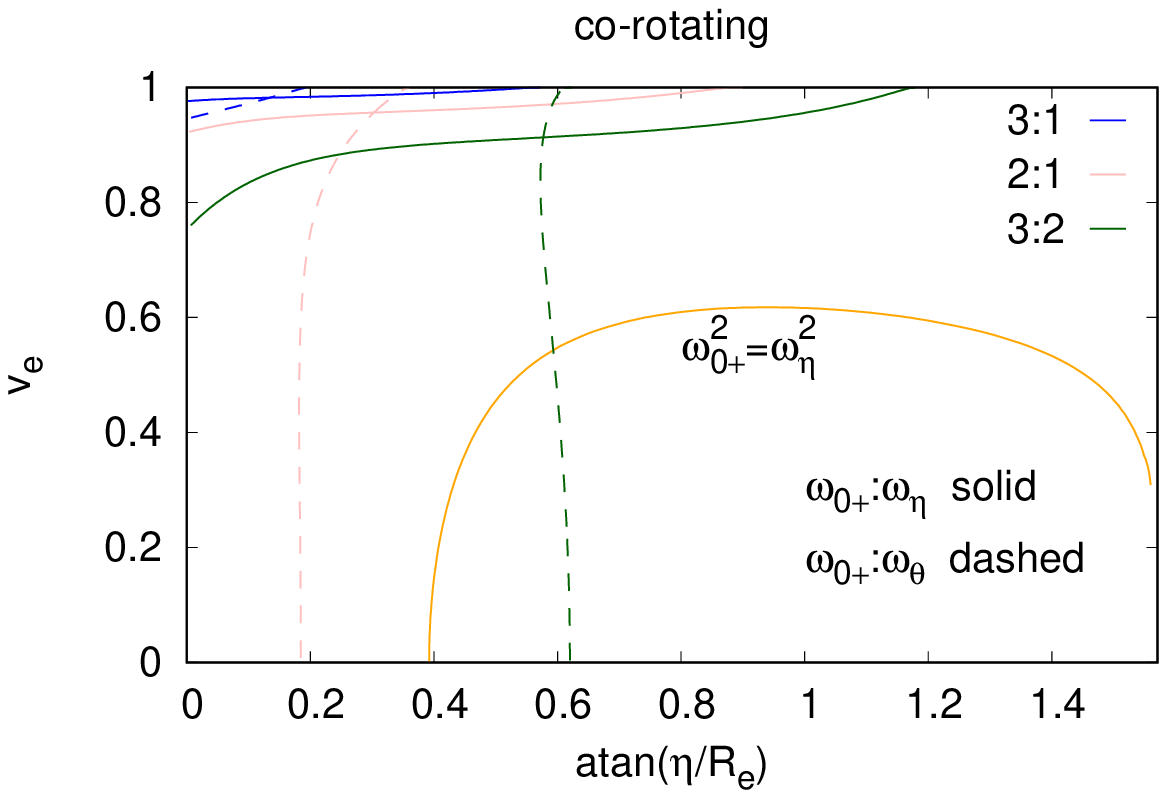}
\includegraphics[height=.24\textheight, angle =0]{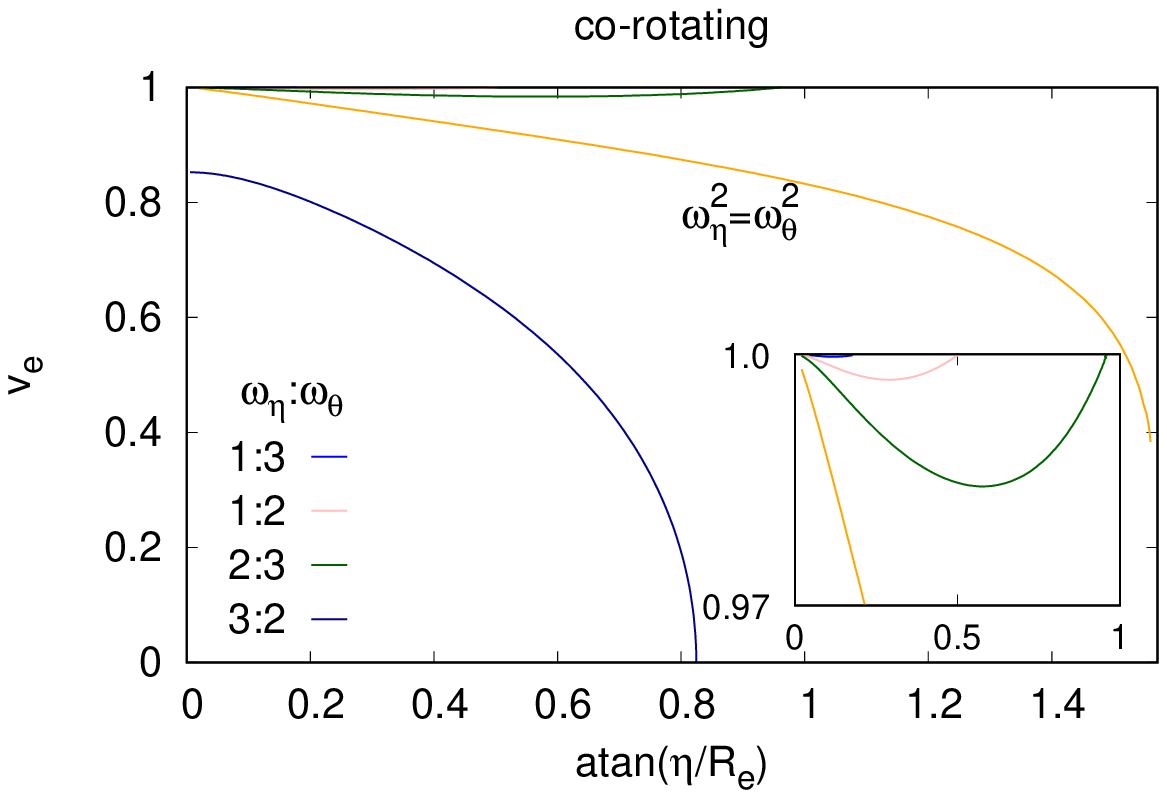}
}
\end{center}
\vspace{-0.5cm}
\caption{
Location of the resonances for co-rotating orbits in the $v_e-\atan (\eta/R_e)$ plane:
(a) Keplerian resonances with radial epicyclic frequencies (solid), and
Keplerian resonances with vertical epicyclic frequencies (dashed).
(b) Parametric and forced resonances
with zoom into the vicinity of the extremal Kerr solution. The resonances with
$\omega_\eta:\omega_\theta = 1:3$ reside only very close to the wormhole throat and to the axis $v_e=1$.
}
\label{fig9}
\end{figure}

We exhibit the location of the lowest resonances in the $v_e-\atan (\eta/R_e)$ plane 
in Fig.\ref{fig9} for co-rotating orbits 
and in Fig.\ref{res_count} for counter-rotating orbits. We restrict ourselves to the values $m, n\leq 3$ since these resonances are supposed to be the strongest and physically most significant. Fig.\ref{fig9}(a) shows the Keplerian resonances with radial epicyclic frequencies (solid),
and the Keplerian resonances with vertical epicyclic frequencies (dashed),
while
Fig.\ref{fig9}(b) exhibits the parametric and forced resonances, 
and contains a zoom into the vicinity of the extremal Kerr solution.  Since we have three regions with different frequency ordering (see Fig.\ref{fig6}), we can expect the excitation of different types of resonances in each of them.  In the uppermost region with very fast rotation we observe all the resonances which are possible for the Kerr black hole. This behavior is expected since for the rotation velocity $\nu_e =1$ we should reproduce the resonance structure of the extremal Kerr solution in a continuous way. Decreasing the rotation velocity so that we cross the curve $\omega_\eta = \omega_\theta$, the parametric and forced resonances for the Kerr solution become impossible. In general the frequency ordering now allows for the resonances  $\omega_\eta: \omega_\theta = 2:1$ and $\omega_\eta: \omega_\theta = 3:1$, but it can be shown that for co-rotating wormholes it is satisfied that $\omega_\eta:\omega_\theta \leq 1.8$. Thus, the lowest possible forced resonance in this region is  $\omega_\eta:\omega_\theta =3:2$. Decreasing further the rotation velocity and  receding from the wormhole throat we reach the region with
$\omega_\eta>\omega_{0+}$ enabling new  Keplerian resonances. However, the frequencies satisfy a further inequality $\omega_{0+}: \omega_\eta \geq 0.88$, which prohibits the lowest resonances $\omega_\eta:\omega_{0+} = m:n$ with $m,n\leq 3$, $m>n$. In summary,  the co-rotating wormholes possess a rich resonance structure for high spin allowing for all the resonances, which are excited in the case of the Kerr black hole. However, most of these modes don't exist for slow rotation and very few new types of resonances arise in the low spin region due to further restrictions on the frequencies. 

We can give an estimate about the lowest possible resonances for slowly rotating wormholes by constructing perturbatively an analytical solution in the near static limit and studying its properties. Such an analysis is presented in the Appendix, while in Fig.$\ref{res_sl}$ we illustrate the possible frequency ratios in this limit. We see that the general bounds $\omega_\eta:\omega_\theta \leq 1.811$ and $\omega_{0+}: \omega_\eta \geq 0.88$ are still valid for slow rotation and we get an additional constraint $\omega_{0+}: \omega_\eta\leq 1.27$ in the region where $\omega_{0+}\leq \omega_\eta$ is satisfied. Thus, for slow rotation we can have Keplerian resonances with the radial epicyclic frequency only in the range $\omega_{0+}: \omega_\eta =m:n$, where $m,n\geq 4$, $m\geq n$ or in the range $m,n \geq 8$, $m< n$.

\begin{figure}[h!]
\begin{center}
(a)\mbox{
\includegraphics[width=.45\textwidth, angle =0]{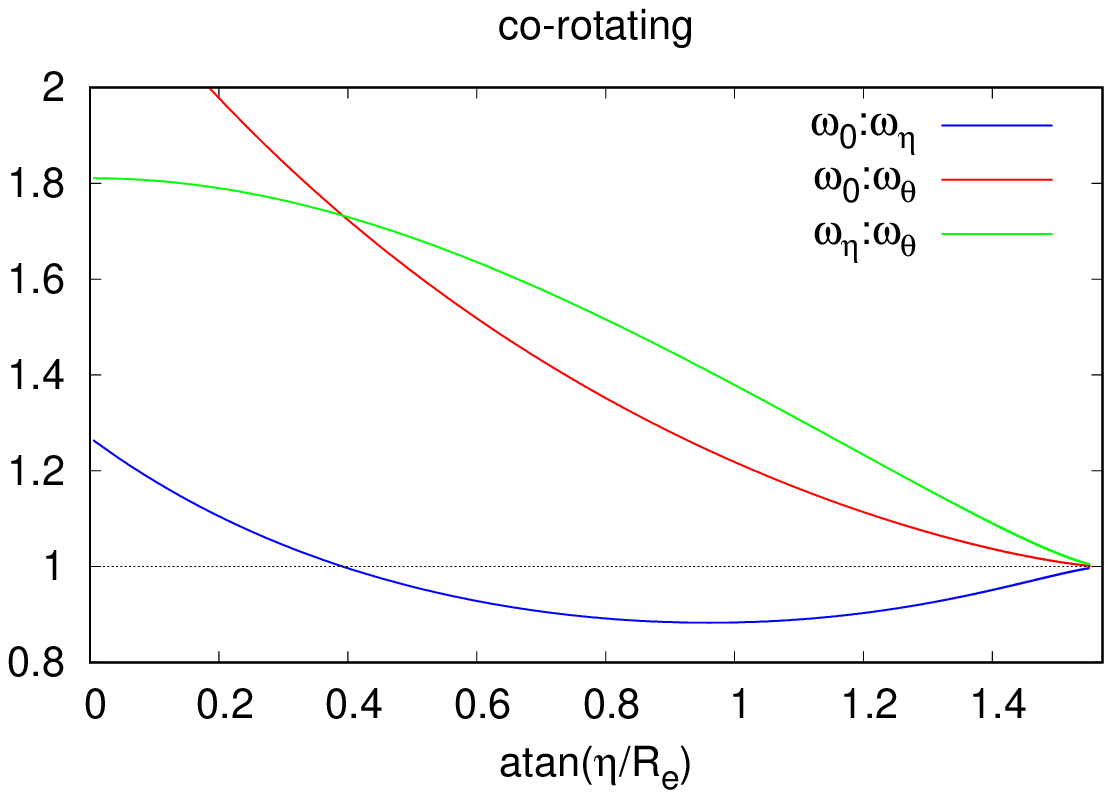} 
(b)      
\includegraphics[width=.45\textwidth, angle =0]{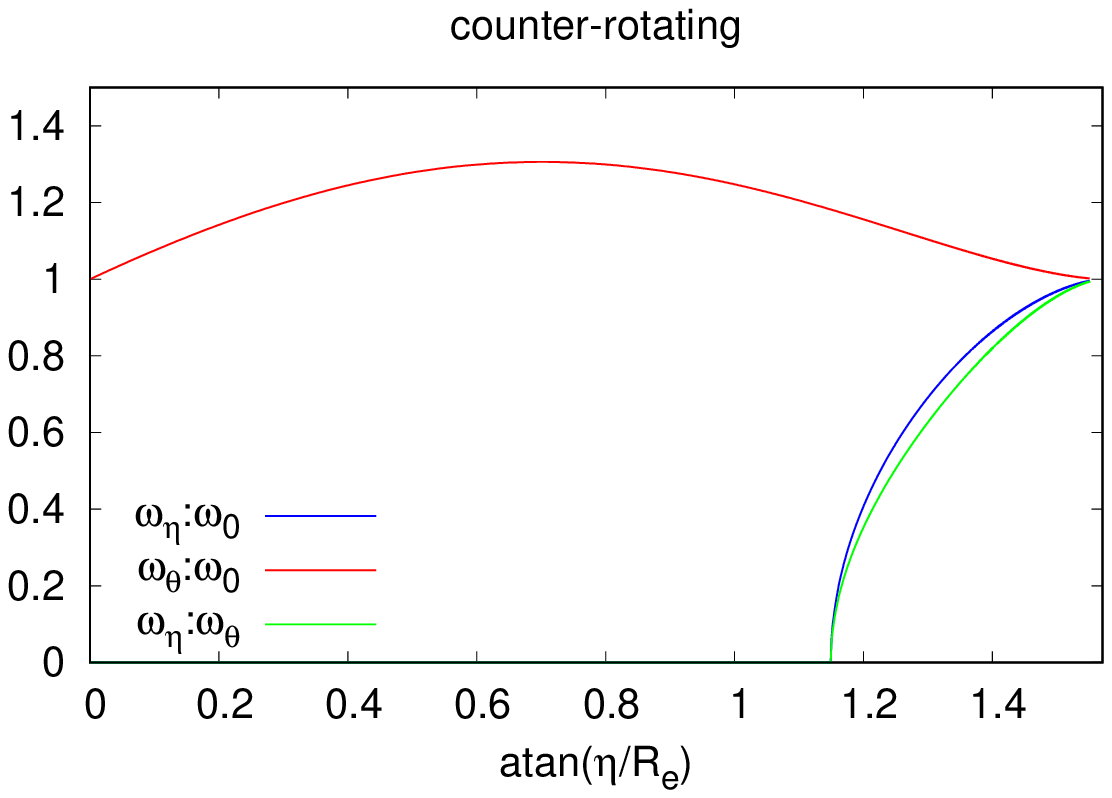}
}
\end{center}
\vspace{-0.5cm}
\caption{
Possible epicyclic and Keplerian frequency ratios in the limit of slowly spinning wormholes for  co-rotating (a), and counter-rotating (b) orbits. }
\label{res_sl}
\end{figure}

\begin{figure}[h!]
\begin{center}
\mbox{
\includegraphics[height=.24\textheight, angle =0]{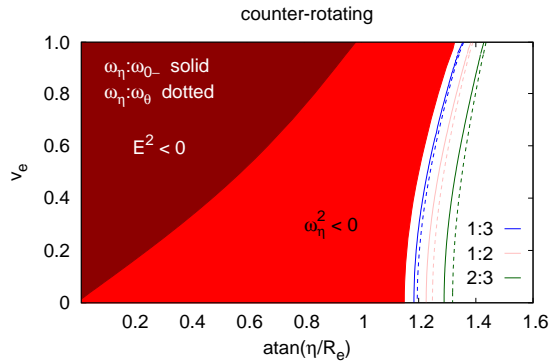}
}
\end{center}
\vspace{-0.5cm}
\caption{
Location of the resonances for counter-rotating orbits in the $v_e-\atan (\eta/R_e)$ plane:
Keplerian resonances with radial epicyclic frequencies (solid),
parametric and forced resonances (dotted).
}
\label{res_count}
\end{figure}

In Fig.\ref{res_count}  we present the lowest resonances for the counter-rotating wormholes. This case resembles closely the Kerr black hole, having the same frequency ordering throughout the parametric space and the same types of possible resonances. 
Thus, we have Keplerian resonances with the radial epicyclic frequency $\omega_\eta:\omega_{0-}$, and parametric and forced resonances, $\omega_\eta:\omega_\theta$,
with frequency ratios $1:3$, $1:2$ and $2:3$. The Keplerian resonances with the vertical epicyclic frequency $\omega_\theta:\omega_{0-}$ are constrained by the property $\omega_{0-}:\omega_\theta > 2/3$, which is satisfied in the domain of existence of the circular orbits $E^2>0$. Therefore, the lowest resonances $\omega_\theta:\omega_{0-}=m:n$ are not allowed and they start first to be excited for integers $m,n \geq 8$, $m>n$.

We now compare the described resonance structure with the Kerr case and  the previously studied family of rotating
Teo wormholes \cite{Deligianni:2021ecz}.
We notice that for the counter-rotating orbits both wormhole solutions behave qualitatively like the Kerr black hole in terms of frequency ordering and types of excited resonances. For co-rotating orbits we get a more complicated picture. Both wormholes contain  Kerr-like regions in their parametric space but they are localized in a different way. For the Teo wormhole this is the region of slowly spinning solutions making a smooth transition through the static limit to the counter-rotating case. When we increase the spin, the Kerr-like behavior is preserved only in a narrow strip around the wormhole throat. Thus, for any spin the same types of co-rotating resonances can be excited as for the Kerr solution, however they are localized in a very close neighbourhood of the throat. The Ellis wormhole demonstrates the opposite behavior. The co-rotating orbits possess Kerr-like properties for rapidly rotating solutions near the extremal value of the spin for the Kerr black hole and deviate from the Kerr case when decreasing the spin. The static limit is singular since it allows no circular orbits outside the wormhole throat (see the Appendix), so there is no continuous transition in the properties of the epicyclic motion to the counter-rotating case. Still there is an important similarity between the two wormhole spacetimes. For fast rotation both allow the same  parametric and forced  co-rotating resonances as for the Kerr solution, however they are excited very closely to the wormhole throat where the epicyclic frequencies reach their highest values. This property is beneficial for  modelling of observational data since these resonances are supposed to lead to strong signals. For example, the location of the parametric resonance $\omega_\theta:\omega_\eta =3:2$ for the Kerr solution is not very satisfactory since for rapidly rotating black holes it resides at comparatively large distance from the ISCO at more than 20 gravitational radii $r_g= GM/c^2$.

Let's consider now the regions in the parametric space where the co-rotating resonances deviate from the properties of the Kerr solution. Both the Ellis and Teo wormholes allow for the same novel frequency orderings $\omega^2_\eta \geq \omega^2_{0+} \geq  \omega^2_\theta$ and $\omega^2_{0+} \geq \omega^2_\eta  \geq \omega^2_\theta$, which enable the excitation of the new types of resonances. However, in the case of the Ellis wormhole some of these possibilities are not realized due to further restriction on the properties of the orbital and epicyclic frequencies. Thus, both wormhole spacetimes allow for lower order parametric resonances than the Kerr solution, for  which the first parametric resonance is excited at $n=3$. However, the Teo wormhole supports all the possible modes with $n=1$ and $n=2$ \cite{Deligianni:2021ecz}, while for the Ellis wormhole we have only the second one since the $n=1$ resonance is prohibited. In a similar way in the Teo's spacetime we can have all the lowest forced resonances enabled by the new frequency ordering $\omega_\eta\geq\omega_\theta$, but for the Ellis wormhole $\omega_\eta:\omega_\theta =2:1$ and $\omega_\eta:\omega_\theta =3:1$ are not allowed. Both wormhole spacetimes contain the Keplerian resonance $\omega_\eta = \omega_{0+}$ but don't allow further low order resonances $\omega_\eta :\omega_{0+} =m:n$, $m, n\geq 3$, $m>n$ in the regions where the frequency ordering deviates from Kerr, i.e. where $\omega_\eta > \omega_{0+}$.

\begin{figure}[h!]
\begin{center}
\mbox{
\includegraphics[height=.24\textheight, angle =0]{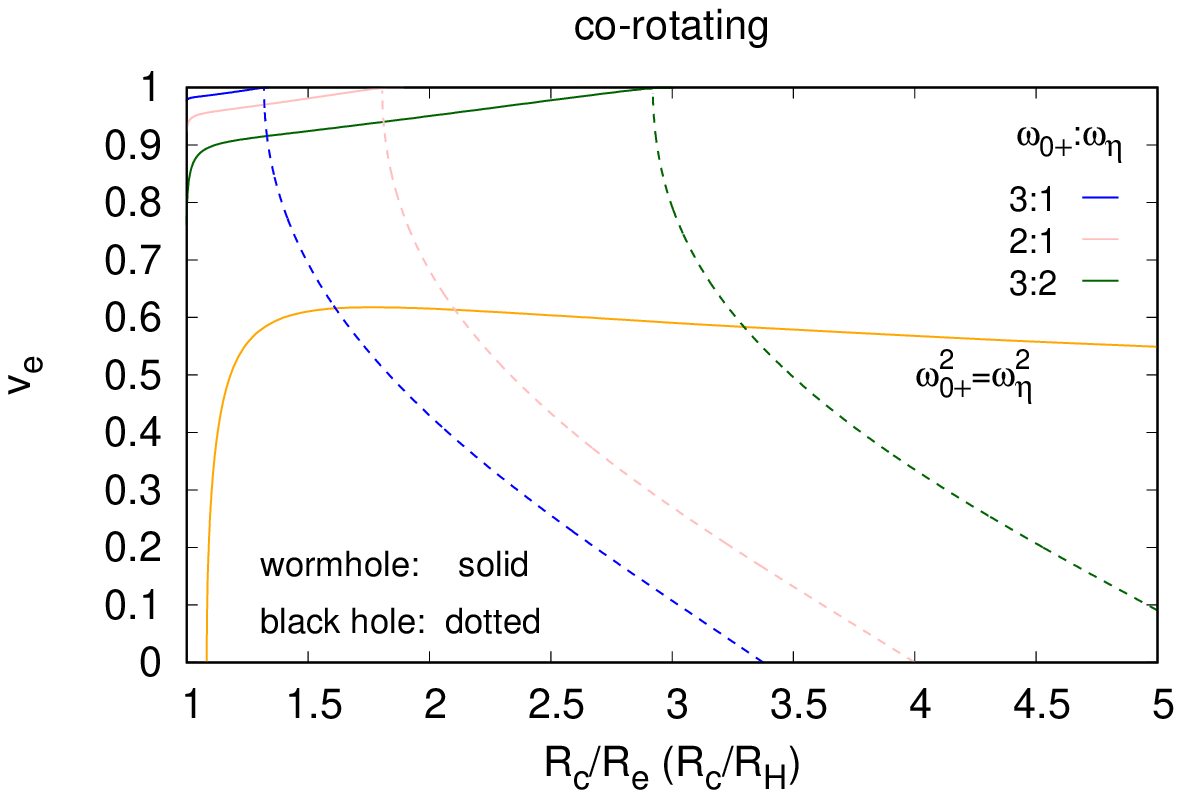}
\includegraphics[height=.24\textheight, angle =0]{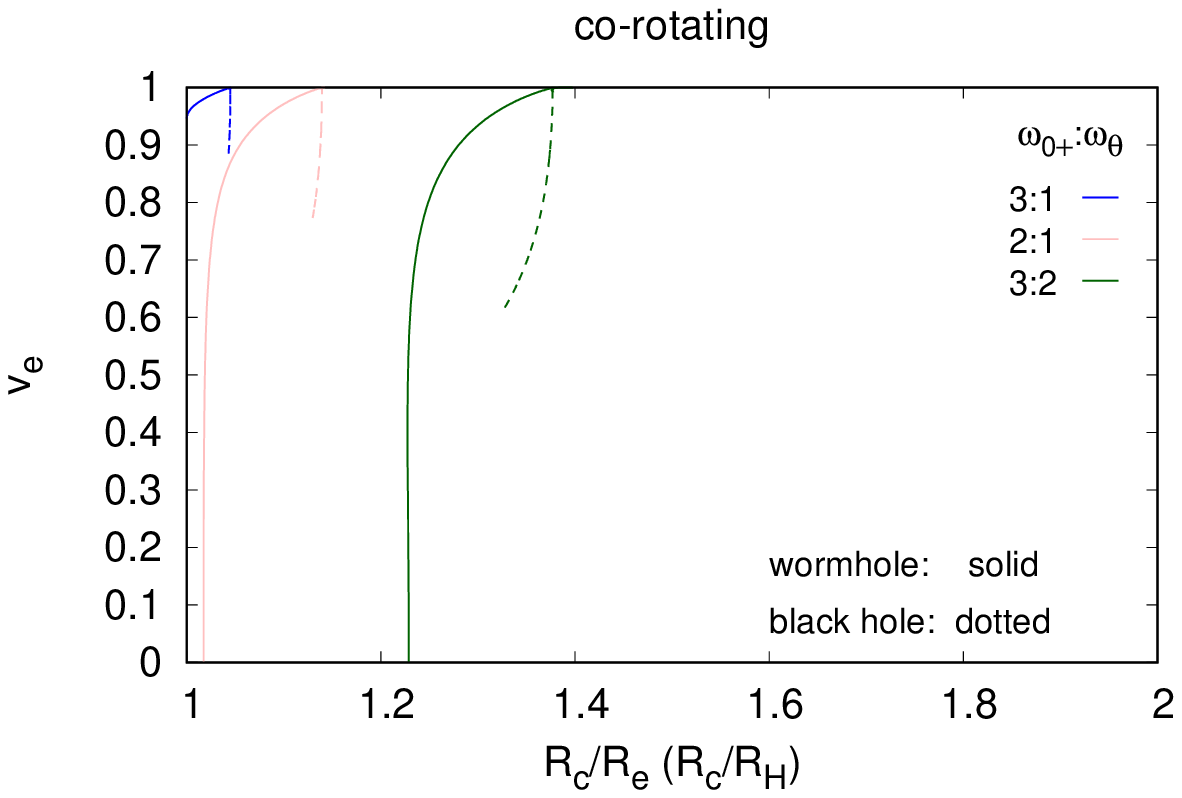}
}
\end{center}
\vspace{-0.5cm}
\caption{
Comparison of the location of the co-rotating Keplerian resonances with the radial  (a), and vertical  (b) epicyclic frequencies for the rotating Ellis wormhole (solid line) and the Kerr black hole (dotted line). Note that the co-rotating resonance lines for the Kerr black hole truncate when the ISCO is reached, i.~e.~when $\omega^2_\eta=0$}.
\label{Ellis_Kerr1}
\end{figure}

\begin{figure}[h!]
\begin{center}
\mbox{
\includegraphics[height=.24\textheight, angle =0]{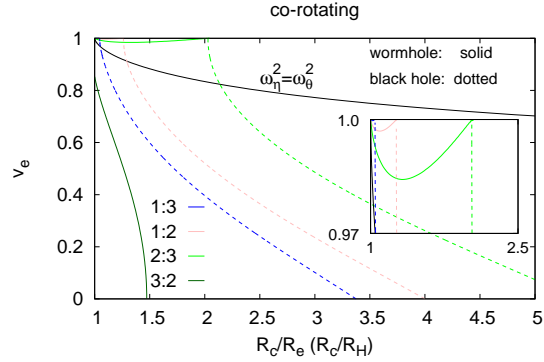}
}
\end{center}
\vspace{-0.5cm}
\caption{
Comparison of the location of the co-rotating parametric and forced resonances  for the rotating Ellis wormhole (solid line) and the Kerr black hole (dotted line).
}
\label{Ellis_Kerr2}
\end{figure}

\begin{figure}[h!]
\begin{center}
\mbox{
\includegraphics[height=.24\textheight, angle =0]{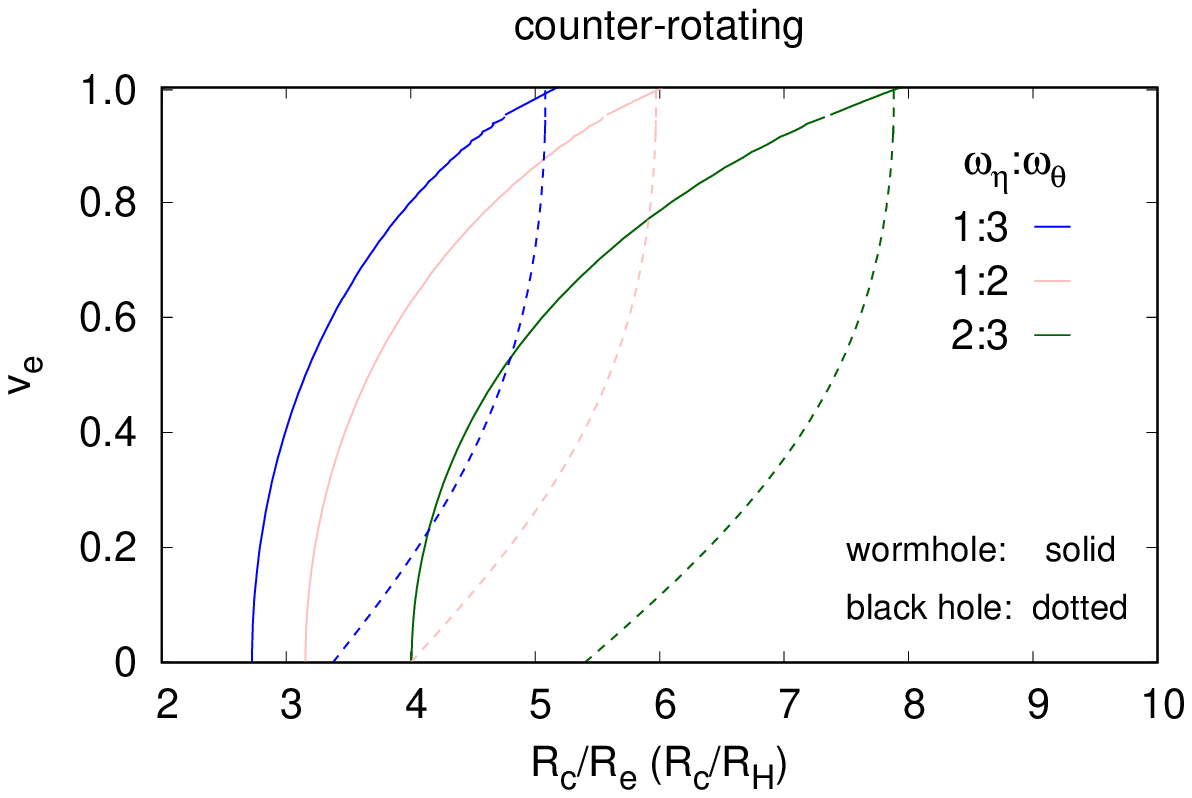}
\includegraphics[height=.24\textheight, angle =0]{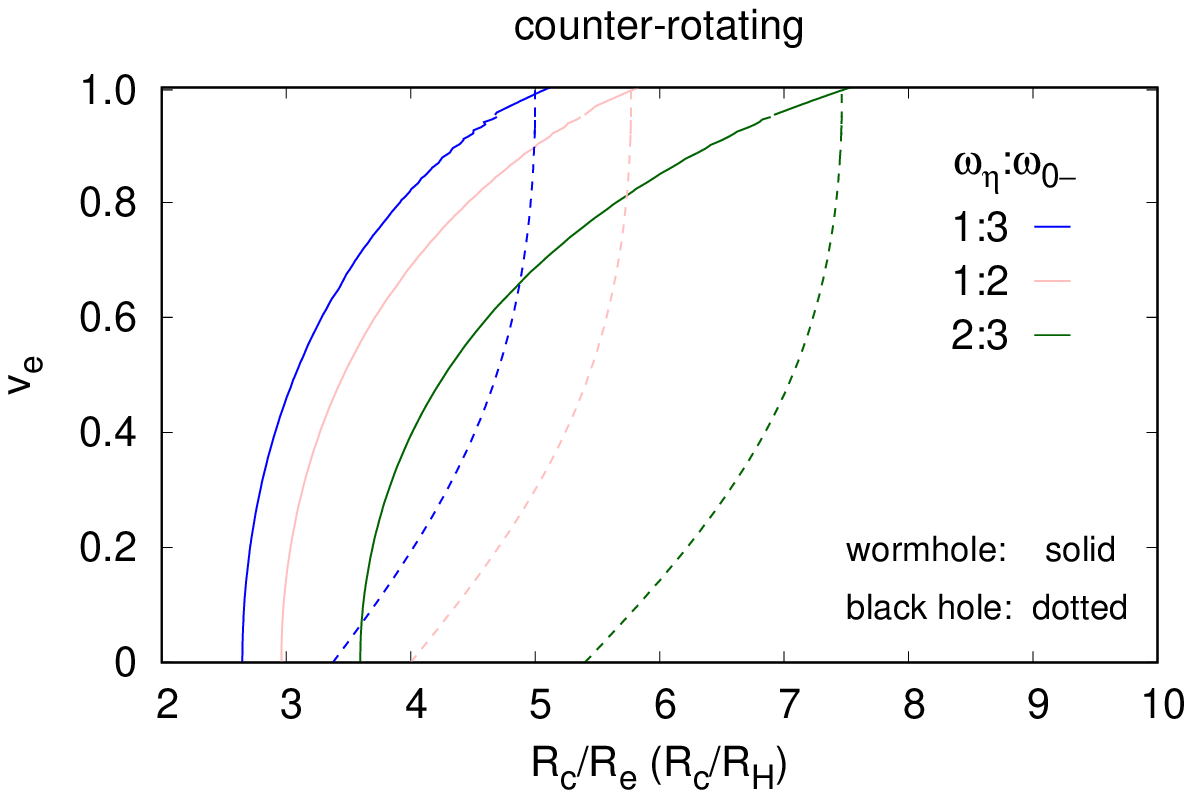}
}
\end{center}
\vspace{-0.5cm}
\caption{
Comparison of the location of the counter-rotating parametric and forced  resonances (a), and the Keplerian resonances  (b)  for the rotating Ellis wormhole (solid line) and the Kerr black hole (dotted line).
}
\label{Ellis_Kerr3}
\end{figure}

In order to gain a more precise understanding of the properties of the Ellis wormholes  in 
Figs.$\ref{Ellis_Kerr1}$-$\ref{Ellis_Kerr3}$ we compare the location of  the resonances with that for the Kerr black hole in the cases when the same types of resonances exist for both spacetimes. Looking at the co-rotating Keplerian resonances in Fig.$\ref{Ellis_Kerr1}$(a) we notice that they are excited only for rapid rotation after a certain critical value of the rotation velocity $\nu^{crit}_e$. When the ratio of the frequencies $\omega_{0+}:\omega_\eta$ increases, the value of $\nu^{crit}_e$ also increases so that the resonance $\omega_{0+}:\omega_\eta= 3:1$  is excited only for velocities approaching the extremal Kerr value $\nu_e=1$. Another characteristic property is that for the critical rotation velocity the resonances reside almost at the throat while when the spin increases they move away. This behavior is opposite to the Kerr black hole, for which the resonance location monotonically approaches the horizon when the spin increases reaching its closest position in the extremal case. However, both for the Kerr black hole and the Ellis wormhole the whole region in which a particular resonance resides moves inwards to smaller gravitational radii when the ratio of the frequencies $\omega_{0+}:\omega_\eta$ increases. Thus, the resonances  $\omega_{0+}:\omega_\eta= 3:1$, for example, are always located closer to the compact object than the $\omega_{0+}:\omega_\eta= 2:1$ resonances.  

The described features are observed quite consistently also for the co-rotating Keplerian resonances with the vertical frequency in Fig.$\ref{Ellis_Kerr1}$(b) and the parametric and forced resonances in Fig.$\ref{Ellis_Kerr2}$, although there are certain exceptions. In some cases the vertical Keplerian resonances can exist for any rotation rate so we don't have always a threshold value of the rotation velocity. The parametric and forced resonances on the other hand don't lead to a monotonic curve in the parameter space but have a local minimum. Thus, in this case the critical value of the rotation velocity after which they are excited corresponds to the local minimum.  For every rotation velocity larger than $\nu^{crit}_e$ the same type of resonances are excited at two different locations one of which approaching the wormhole throat.

Next we consider another interesting feature of the Ellis wormhole by studying the possibility for excitation of triple resonances. Triple resonances arise  when the epicyclic  and the orbital frequencies obey the ratio $\omega_0:\omega_\theta:\omega_\eta =m:n:k$ at a certain point of the parametric space, where $m$, $n$ and $k$ are small positive integers. These resonances are expected to be stronger since the individual resonances between each couple of frequencies are supposed to be causally connected when excited at the same radius and amplify each other. They were further motivated by the observation of more than two high-frequency oscillations in some X-ray sources. For example, three high frequency peaks are reported for Sgr A* and NGC 5408 X-1  with frequency ratios 3:2:1 and 6:4:3, respectively \cite{Aschenbach:2004}-\cite{Strohmayer:2007}. 

\begin{figure}[h!]
\begin{center}
\mbox{
\includegraphics[height=.24\textheight, angle =0]{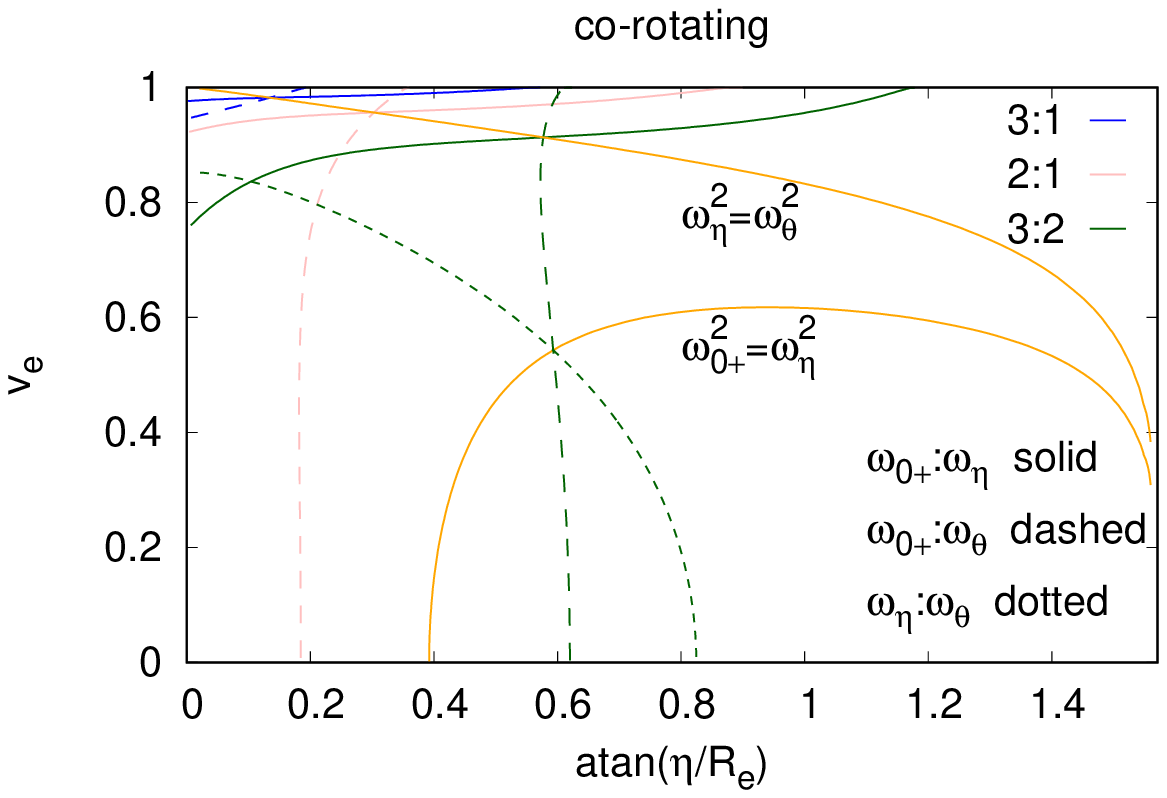}
\includegraphics[height=.24\textheight, angle =0]{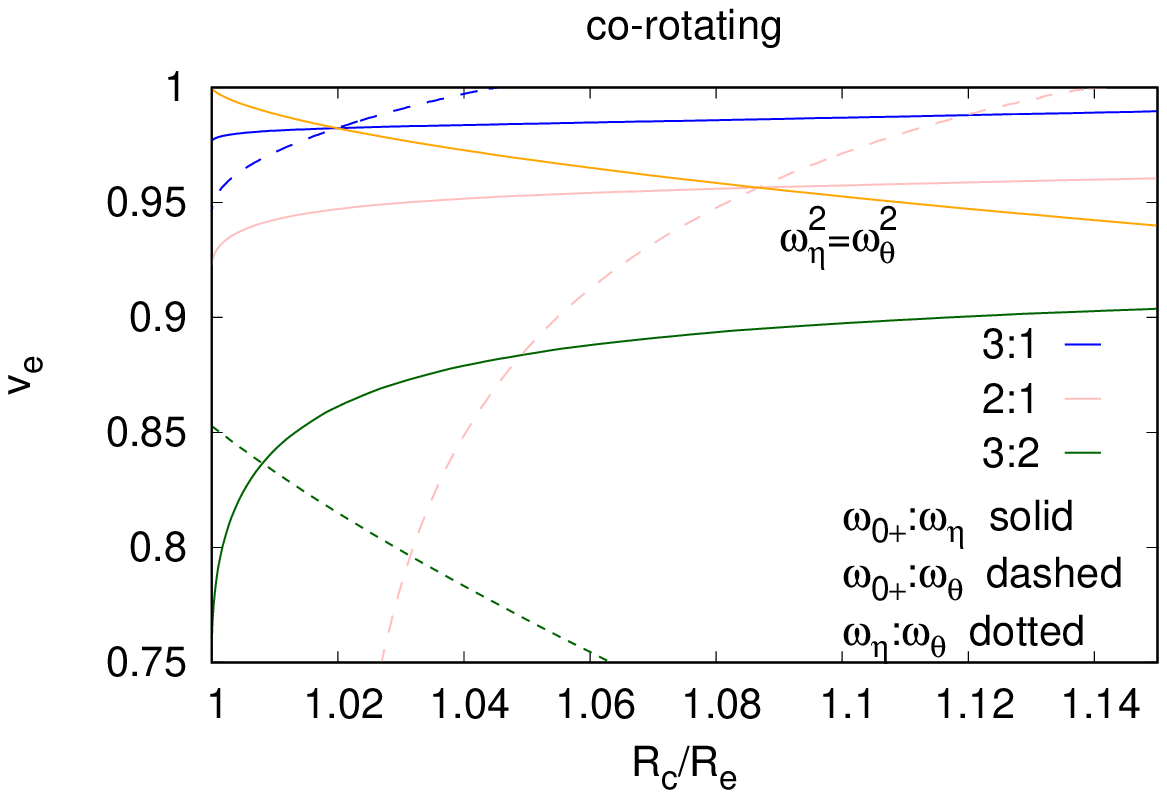}
}
\end{center}
\vspace{-0.5cm}
\caption{
Location of the triple resonances. In the  right panel we show a zoom of the vicinity of the wormhole throat for rapid rotation.
}
\label{triple_res}
\end{figure}

Considering the Kerr black hole we see that this phenomenon is rare. For sufficiently low integers $m,n,k \leq 5$  only a single triple resonance can be excited in this case. It is characterized by the frequency ratio $\omega_0:\omega_\theta:\omega_\eta =3:2:1$ and occurs at the radial coordinate $r/M = 2.395$ for nearly extremal spin $a/M \approx 0.983$. For wormhole spacetimes, however, we can have more instances. In Table 1 we list the possible triple resonances for the Ellis wormhole with frequency ratios $\omega_0:\omega_\theta:\omega_\eta =m:n:k$, $m,n,k \leq 6$, where we also give their radial position and the characteristic rotational velocity of the wormhole throat. They  are further visualized in Fig.$\ref{triple_res}$ corresponding to the different intersections between the resonance curves. We note that  the triple resonances  $\omega_0:\omega_\theta:\omega_\eta =4:3:2\,$; $4:2:1\,$; $6:2:1$ and $3:2:1$ are not presented in the figures since they result from coinciding locations of the Keplerian and parametric or forced resonances  for very high rotational velocity and very near the wormhole throat.

\begin{table}[h!]
\centering
\begin{tabular}{ |p{1.5cm}||p{1.0cm}|p{1.0cm}|p{1.0cm}|p{1.0cm}|p{1.0cm}|p{1.0cm}|p{1.0cm}|p{1.0cm}|p{1.0cm}|p{1.0cm}|p{1.0cm}|}
 \hline
 \multicolumn{12}{|c|}{Triple resonances} \\
 \hline
 $\omega_{0+}$:$\omega_\theta$:$\omega_\eta$ & 2:1:1   & 3:1:1   & 3:2:2   & 3:2:3   & 4:2:1    & 4:2:3   & 4:3:2   &  6:2:1   & 6:3:2   & 6:3:4   & 6:4:3  \\
 \hline
 $R_c/R_e$                                   & $1.087$ & $1.02$  & $1.29$  & $1.23$  &  $1.13$  & $1.03$  &  $1.61$ & $1.04$   & $1.12$  & $1.05$  & $1.33$ \\
 \hline
$v_e$                                        & $0.956$ & $0.982$ & $0.913$ & $0.543$ &  $0.997$ & $0.796$ & $0.988$ & $0.998$  & $0.988$ & $0.883$ & $0.971$ \\
 \hline
\end{tabular}
\caption{Possible triple resonances for the rotating Ellis wormhole. For each resonance we give the ratio of the epicyclic and orbital frequencies, the radial position $R_c/R_e$ and the corresponding  rotation velocity of the wormhole throat $\nu_e$.}
\label{Tab:1}
\end{table}

\section{Conclusions}

Recent astrophysical experiments, which  probe the gravitational field in the regime of strong interaction, will expand our knowledge about the properties of  black holes. However, they could further bring insights about some more exotic compact objects like wormholes or naked singularities. Wormholes often mimic black holes in their phenomenological behavior. Therefore, it is important to consider them as a black hole alternative when interpreting the experimental data.

An interesting question is how closely the X-ray spectrum from the accretion disks around wormholes can resemble the case of black holes, or how precisely we can differentiate between these two types of compact objects by means of X-ray spectroscopy. In this regard we study the quasi-periodic oscillations from the accretion disk around rotating Ellis wormholes. We interpret them in the framework of the resonance models where the QPOs correspond to particular non-linear resonances between the epicyclic and orbital frequencies of the quasi-circular motion in the equatorial plane. Then, we compare the resonance structure of the rotating Ellis spacetime with another rotating wormhole geometry, which we studied recently, and try to find some common features, which can  distinguish a wormhole spacetime from the Kerr black hole.

We observe that both wormhole spacetimes contain regions in the parametric space where the properties of the epicyclic motion and the possible resonances resemble the Kerr black hole. For counter-rotating orbits the same types of resonances can be excited in wormhole and black hole spacetimes and they are located at a similar radial distance from the compact object. For co-rotating orbits, however, the wormhole solutions are characterized  by a very diverse resonance structure which depends on the wormhole spin and includes new types of resonances  in certain regions. A common feature is that they both allow parametric and forced resonances of lower order than for the Kerr black hole, which are expected to be stronger. In addition, a wide range of resonances are excited in the region with very strong gravitational field in the vicinity of the wormhole throat. Another interesting property of the rotating Ellis spacetime is that it allows for a number of triple resonances of sufficiently low order, while for the Kerr black hole only a single such case is possible. These resonances arise due to the coupling between all the three characteristic frequencies at the same radial distance and could explain the evidence for more  than two high-frequency peaks in the observed spectrum from some X-ray sources.  

\section{Appendix}

\subsection{Alternative expressions for the frequencies}

Here we present alternative expressions for the squares of the frequencies 
\begin{eqnarray}
\omega_0^2 & = & \frac{(e^{2f} l - h l \omega^2 + h \omega)^2}{(l \omega - 1)^2 h^2} \ ,
\label{om0al}\\
\omega_{\eta}^2 & = &  
\frac{
\left[2 (4 \eta^2 - h) + f_{,\eta\eta} h^2 + (f_{,\eta} h - 4 \eta) f_{,\eta} h\right] 
e^{4f} l^2
}{2 (l \omega - 1)^2 e^{\nu} h^3} 
\nonumber\\
& & -
\frac{
\left[((f_{,\eta}^2 - f_{,\eta\eta}) (l \omega - 1) - 4 f_{,\eta} l \omega_{,\eta}) (l \omega - 1) 
 + 2 ((\omega \omega_{,\eta\eta} +
 \omega_{,\eta}^2) l - \omega_{,\eta\eta}) l\right]  e^{2f}
}{2 (l \omega - 1)^2 e^{\nu}} \ ,
\label{omeal}\\
\omega_\theta^2 & = & \frac{(((l \omega - 1) f_{,\theta\theta} - 2 l \omega_{,\theta\theta}) (l \omega - 1) h + (f_{,\theta\theta} + 2) e^{2f} l^2) e^{2f}
}{2 (l \omega - 1)^2 e^{\nu} h^2} \ ,
\label{omtal}
\end{eqnarray}
where the proper angular momentum  $l=L/E$ is a solution of the quadratic equation
\begin{equation}
l^2 - 2 l
\frac{ h^2 (\omega f_{,\eta}-\omega_{,\eta} )}{(h f_{,\eta}-2 \eta) e^{2f}+(\omega f_{,\eta} - 2\omega_{,\eta}) h^2\omega}
+ \frac{h^2 f_{,\eta} }{(h f_{,\eta}-2 \eta) e^{2f}+(\omega f_{,\eta} - 2\omega_{,\eta}) h^2\omega}
=0 \ ,
\label{eql2}
\end{equation}
and
\begin{equation}
E^2 = \frac{e^f h}{(l\omega - 1)^2 h - e^{2f} l^2} \ .
\label{eqe2}
\end{equation}

\subsection{Special cases}

Let us now consider the frequencies for two special cases, (i) the static massless Ellis wormhole
and (ii) the throat of the rotating Ellis wormhole.
We note, that the epicyclic frequencies of static wormholes
have been considered recently
also in a more general setting
\cite{DeFalco:2021btn}.

(i) For the static massless Ellis wormhole the solution is known in
closed form,
\begin{equation}
f=0 \ , \ \ \ 
\nu = 0 \ , \ \ \ 
\omega = 0 \ .
\label{statwh}
\end{equation}
Since the energy $E$ and proper angular momentum $l$ of a particle are related to 
$\dot{t}$ and $\dot{\varphi}$ by 
\begin{equation}
\dot{t} = E \ , \ \ \ \dot{\varphi}= E l/h \ ,
\label{tpandphip}
\end{equation}
we find for the orbital frequency 
$ \omega_0 = l/h $. 
Substitution of the solution Eq.~(\ref{statwh})
into $l^2$ given in (\ref{eql2}) shows, that for $\eta\ne 0$ always $l=0$,
thus away from the throat the proper angular momentum of the particle
always vanishes.
In contrast, at the throat $l$ may be finite, since no condition arises from (\ref{eql2}).
Substitution of the solution Eq.~(\ref{statwh})
into $\omega^2_\eta$ and $\omega^2_\theta$, given in (\ref{omeal})
and (\ref{omtal}), leads to
\begin{eqnarray}
\omega^2_\eta & = & \omega^2_0 \left(-1+4\frac{\eta^2}{h}\right) =
-\omega^2_0 +4 \frac{\eta^2 l^2}{h^3}
\ , 
\label{o2x_static} \\
\omega^2_\theta & = & \omega^2_0 \ . 
\label{o2y_static}
\end{eqnarray}
On the other hand, circular orbits are determined by the conditions
\begin{equation}
0=V_{\rm eff} (\eta)= E^2 -1 - E^2 \frac{l^2}{h} \ , \ \ \ 
0=V_{{\rm eff},\eta} (\eta)=  2 E^2  \frac{\eta l^2}{h^2} \ .
\label{orbcon_stwh} 
\end{equation}

To interpret these results, we need to consider the throat $\eta=0$ separately.
Let us first consider a radial coordinate outside the throat. 
Then the angular momentum of the particle must vanish, $l=0$.
Thus a particle may sit at rest anywhere in the spacetime.
However, all frequencies vanish as well,
$\omega^2_0=\omega^2_\eta=\omega^2_\theta =0$,
i.e., all points outside the throat represent marginally stable points.
At the throat $\eta=0$, however, we may allow for $l \ne 0$.
Thus orbits are possible, but since the above conditions yield
$\omega^2_\eta = -\omega^2_0$,
these orbits are not stable.

\begin{figure}[t!]
\begin{center}
(a)\mbox{
\includegraphics[width=.40\textwidth, angle =0]{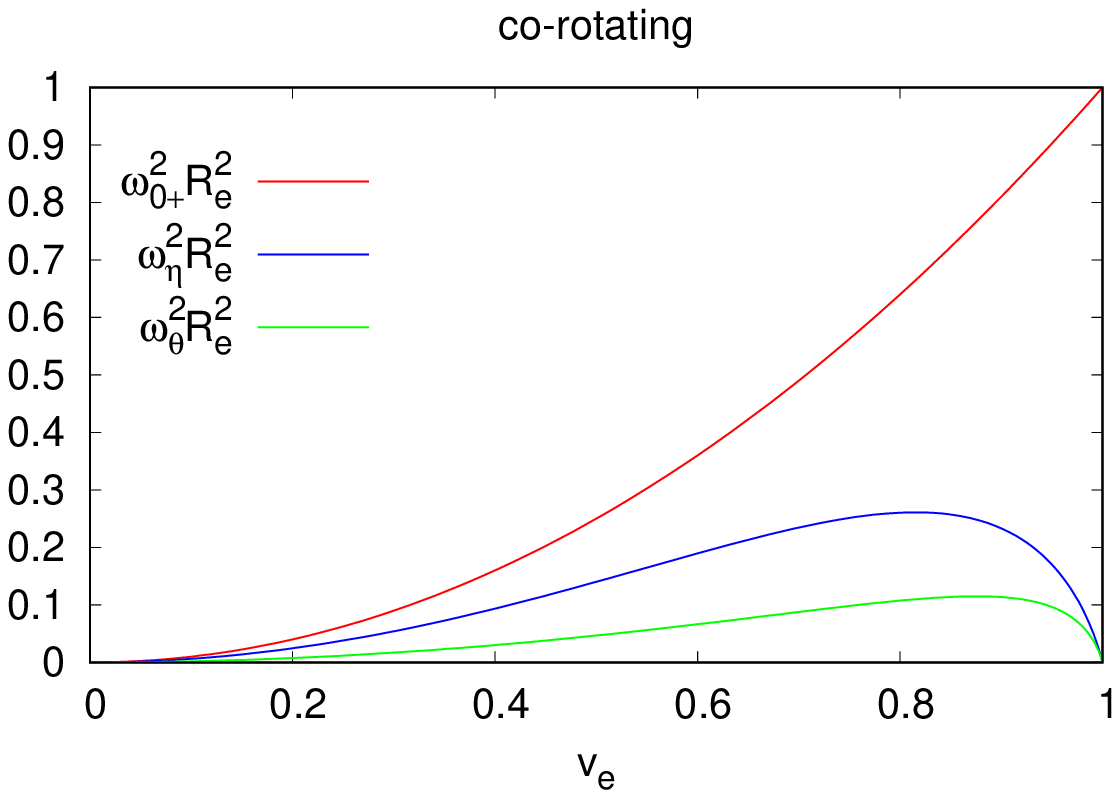} 
(b)      
\includegraphics[width=.40\textwidth, angle =0]{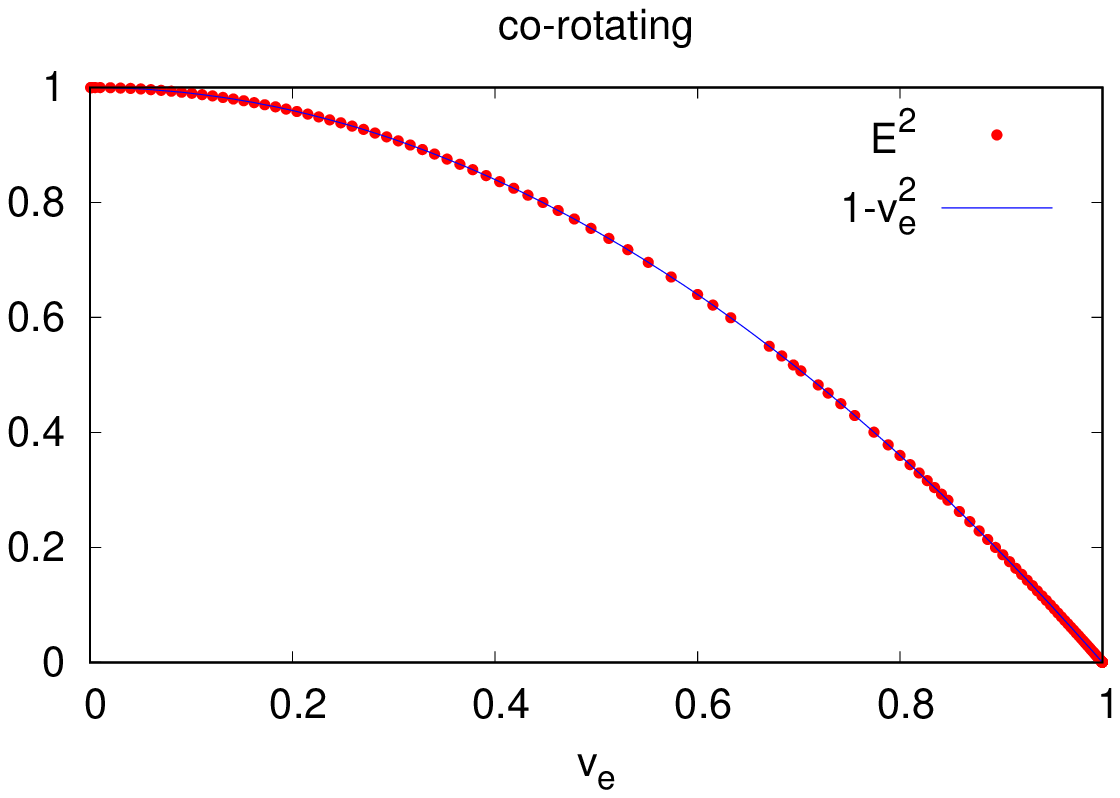}
}
\end{center}
\caption{Throat properties:
(a) scaled squared orbital frequencies $\omega_0^2 R_e^2$,
radial epicyclic frequencies $\omega_\eta^2 R_e^2$,
and vertical epicyclic frequencies $\omega_\theta^2 R_e^2$
vs throat velocity $v_e$;
(b) squared specific energy $E^2$ and quadratic relation $1-v_e^2$
vs throat velocity $v_e$.
\label{app1}
}
\end{figure}

(ii) We now consider the special case where a particle orbits at the throat
of a rotating symmetric wormhole.
We start with an expansion of the metric functions at the throat,
\begin{eqnarray}
f & = & f_0(\theta) + f_2(\theta)\frac{\eta^2}{2} + {\cal O}(\eta^4) \ ,
\label{expthf} \\
\nu & = & \nu_0(\theta) + \nu_2(\theta)\frac{\eta^2}{2} + {\cal O}(\eta^4) \ ,
\label{expthnu} \\
\omega & = & \omega_{th} + \omega_1(\theta) \eta + {\cal O}(\eta^3) \ ,
\label{expthom} 
\end{eqnarray}
where we have taken into account the symmetry properties of the 
metric functions.
Substitution of the expansion into the field equations reveals the relations at 
$\theta=\pi/2$
\begin{equation}
f_2 = -\partial_{\theta\theta}f_0/\eta_0^2
      +e^{2 f_0} \omega_1^2\eta_0^2 \ , 
\ \ \ 
\nu_2 = -\partial_{\theta\theta}\nu_0/\eta_0^2
      +e^{2 f_0} \omega_1^2\eta_0^2 \ . 
\label{relexpthr}
\end{equation}
Expanding the expressions for the proper angular momentum (\ref{eql2}), squared energy 
(\ref{eqe2}) and orbital frequency (\ref{om0al}) yields
\begin{eqnarray}
0 & = & l\omega_1 \left(l \omega_{th} -1\right)  
\ , \label{xpaeql}\\
E^2 & = & \frac{e^{f_0} \eta^2_0}{\left(l \omega_{th} -1\right)^2-e^{2 f_0} l^2}
\ , \label{xpae2}\\
\omega_0  & = & \omega_{th} - \frac{e^{2 f_0} l}{\left(l \omega_{th} -1\right)\eta_0^2}
\ . \label{xpaom0}
\end{eqnarray}
Eq.~(\ref{xpaeql}) possesses two solutions, $l=0$ and $l=1/\omega_{th}$,
where the second solution should be dismissed, 
since it would yield negative $E^2$ and diverging orbital frequency.
From the first solution we find
\begin{eqnarray}
E^2 & = & e^{f_0}\eta^2_0 \ ,
\label{e2_th}\\
\omega_0  & = & \omega_{th} \ , 
\label{om0_th}\\
\omega^2_\eta  & = & \frac{e^{-\nu_0}}{2\eta_0^2}
                     \left(\omega_1^2 \eta_0^4 - e^{2 f_0}\partial_{\theta\theta}f_0\right) \ , 
\label{o2x_th}\\
\omega^2_\theta  & = & \frac{e^{-\nu_0}}{2\eta_0^2} e^{2 f_0}\partial_{\theta\theta}f_0  \ .
\label{o2y_th}
\end{eqnarray}
We note from (\ref{e2_th}) that the particle is dragged along with the same orbital 
frequency as the rotational frequency of the throat.
Adding the squares of the epicyclic frequencies (\ref{o2x_th}) 
and (\ref{o2y_th}) yields a non-negative quantity
\begin{equation}
\omega^2_\eta +  \omega^2_\theta = \omega_1^2 \frac{e^{-\nu_0}}{2}\eta_0^2 \geq 0 \ .
\end{equation}
All three scaled squared frequencies at the throat are shown versus the velocity
$v_e$ of the throat in Fig.\ref{app1}(a).
Clearly, $0\le \omega^2_\theta \le \omega^2_\eta \le \omega_0^2$.
Thus these orbits are stable, except for the static case ($v_e=0$)
and the extremal Kerr limit ($v_e=1$).
Fig.\ref{app1}(b) demonstrates that the relation
$E^2 = 1 - v_e^2$ holds at the throat.
		     
\subsection{Slowly rotating wormholes}

We study the case of slowly rotating wormholes by constructing analytically an approximate solution in this limit (see also \cite{Kashargin:2007mm,Kashargin:2008pk}).
For this purpose we consider the metric given by 
Eq.~($\ref{lineel}$), being interested in wormholes which are symmetric under the interchange of the two asymptotic regions. Then we construct perturbatively rotating solutions by expanding the metric functions in terms of the angular velocity  of the throat, e.g.

\begin{equation}
f = f_0+\lambda f_1 +\lambda^2 f_2 + {\cal O}\left(\lambda^3\right),
\label{expinlam}
\end{equation}
where $\lambda=\omega_{th} = \left. \omega\right|_{\eta=0}$, and similarly for the other metric functions $\nu$, $\omega$ and $\phi$.
In the zeroth order case the solution is known in closed form \cite{Ellis:1979bh, Bronnikov:1973fh}
\begin{equation}
f_0=0\ , \ \ \ \nu_0=0 \ , \ \ \ \omega_0=0 \ , \ \ \ 
\phi_0 = \frac{D}{\eta_0}\left(\atan \,x-\frac{\pi}{2}\right),
\label{ord0}
\end{equation}
where $x=\eta/\eta_0$ and $D$ denotes the scalar charge.

First order perturbation yields
\begin{equation}
\omega = \lambda \omega_1 =\lambda \frac{1}{\pi}
\left(\pi -2\atan\,x -\frac{2x}{1+x^2}\right)
\label{ord1}
\end{equation}
and $f_1=0$, $\nu_1=0$, $\phi_1=0$.

Including the second order terms the metric and phantom field functions read
\begin{eqnarray}
f & = & \lambda^2 F(\eta,\theta) \ ,
\label{ford2}\\
\nu & = & \lambda^2 G(\eta,\theta) \ ,
\label{nuord2}\\
\omega & = & \lambda \omega_1 \ ,
\label{omord2}\\
\phi & =  &\phi_0+\lambda^2 H(\eta,\theta) \ .
\label{phiord2}
\end{eqnarray}
Substitution in the Einstein and scalar field equation and 
considering only the lowest terms in $\lambda$ then yields a set of
linear equations for $F(\eta,\theta)$, $G(\eta,\theta)$, and 
$H(\eta,\theta)$. 
Solutions which vanish in the asymptotic region $\eta \to \infty$
and satisfy the symmetry conditions 
$F_{,\eta} =0$, $G_{,\eta} =0$, 
at the throat $\eta=0$, can easily be found
\begin{eqnarray}
 F(\eta,\theta) & = &
-\frac{\eta_0^2}{\pi^2}
\left[\frac{2\pi^2}{4}+\frac{8}{3(1+x^2)} -\frac{8 \atan^2\, x}{3}
\right. \nonumber\\
& & 
\ \ \ \ \ \ \ \ \ \left. 
+\left(\frac{8(2+3x^2)}{3(x+x^2)}-\frac{2\pi^2(1-3 x^2)}{3}
+16 x \atan\, x +\frac{8(1+3x^2)}{3} \atan^2\, x\right) P_2(\cos\theta)
\right] \ , 
\label{Ford2} \\
 G(\eta,\theta) & = &
-\frac{8}{3 \pi^2 \eta_0^2 (1+x^2)}\left(1-P_2(\cos\theta)\right)\ , 
\label{Gord2} \\
 H(\eta,\theta) & = &
 -\frac{2 D \eta_0}{\pi^2}\left(\atan\, x-\frac{\pi}{2}\right)\ , 
\label{Hord2}
\end{eqnarray}
where $P_2(\cos\theta)$ is the second Legendre polynomial.


Next we consider the expressions for the squares of the frequencies. 
We substitute the perturbed solution in the expressions $\omega_0^2$, $\omega_{\eta}^2$, $\omega_{\theta}^2$, and
in the equation for the proper angular momentum and expand to the lowest order in 
$\lambda$. This yields 
\begin{eqnarray}
\omega_0^2 & = & \lambda^2 \frac{(\hl +  h\omega_1)^2}{h^2} \ ,
\label{pertom0al}\\
\omega_{\eta}^2 & = & 
\lambda^2 \frac{2\left(h^3\omega_{1,\eta\eta} - h\hl + 4 \eta^2\hl\right)\hl +h^3 F_{,\eta\eta}}{2h^3} \ ,
\label{pertomeal}\\
\omega_{\theta}^2 & = &\lambda^2  \frac{h F_{,\theta\theta} + 2\hl^2}{2h^2} \ ,
\label{pertomtal}
\end{eqnarray}
where 
\begin{equation}
l = \frac{h}{2\eta}\left(h \omega_{1,\eta}\pm W\right)
\label{l_pm}
\end{equation}
for the co-rotating, resp. counter-roting case, and $W = \sqrt{2 \eta F_{,\eta} +h(\omega_{1,\eta})^2}$.

Clearly, all frequencies vanish as $\lambda \to 0$. However, the ratios are finite.
We show the ratios $\omega_0:\omega_\eta$,  $\omega_0:\omega_\theta$ and $\omega_\eta:\omega_\theta$ for the
co-rotating case in 
Fig.$\ref{res_sl}$(a) and $\omega_\eta:\omega_0$,  $\omega_\theta:\omega_0$ and $\omega_\eta:\omega_\theta$ for the
counter-rotating case in 
Fig.$\ref{res_sl}$(b).
Note that $\omega^2_\eta > 0 $ in the co-rotating case, but 
 $\omega^2_\eta < 0 $ for $\atan(\eta/R_e)< 1.15$ in the counter-rotating case.

\noindent{\textbf{~~~Acknowledgement.--~}}
We gratefully acknowledge support by the DFG Research Training Group 1620 ``Models of Gravity''  and the COST Actions CA16214 and CA16104.  P.N. is partially supported by the Bulgarian NSF Grant  KP-06-H38/2.

\end{document}